\documentclass[twocolumn]{aastex701}

\submitjournal{ApJ}
\begin{document}

\title{Evolutionary pathways toward survival of a thick CO$_2$- or SO$_2$-rich \\atmosphere on the lava world TOI-561\,b}

\author[0009-0009-5036-3049]{Emma Postolec}
\affiliation{Kapteyn Astronomical Institute, University of Groningen; Groningen, The Netherlands}
\email{e.n.postolec@rug.nl}

\author[0000-0002-3286-7683]{Tim Lichtenberg}
\affiliation{Kapteyn Astronomical Institute, University of Groningen; Groningen, The Netherlands}
\email{tim.lichtenberg@rug.nl}

\author[0009-0008-2801-5040]{Johanna K. Teske}
\affiliation{Earth and Planets Laboratory, Carnegie Institution for Science, 5241 Broad Branch Road, NW, Washington, DC 20015, USA}
\affiliation{The Observatories of the Carnegie Institution for Science, 813 Santa Barbara St., Pasadena, CA 91101, USA}
\email{jteske@carnegiescience.edu}

\author[orcid=0000-0002-8368-4641]{Harrison Nicholls}
\affiliation{Institute of Astronomy, University of Cambridge, Cambridge, CB3 0HA, United Kingdom}
\email{harrison.nicholls@ast.cam.ac.uk}

\author[0000-0002-7971-7439]{Mara Attia}
\affiliation{Kapteyn Astronomical Institute, University of Groningen; Groningen, The Netherlands}
\email{attia@astro.rug.nl}

\author[0000-0002-4487-5533]{Anjali Piette}
\affiliation{School of Physics \& Astronomy, University of Birmingham, Edgbaston, Birmingham, B15 2TT, UK}
\email{a.a.a.piette@bham.ac.uk}

\author[0000-0003-4987-6591]{Lisa Dang}
\affiliation{Waterloo Centre for Astrophysics and Department of Physics and Astronomy, University of Waterloo; Waterloo, Ontario, Canada N2L 3G1}
\email{lisa.dang@uwaterloo.ca}

\author[0000-0003-0354-0187]{Nicole L. Wallack}
\affiliation{Earth and Planets Laboratory, Carnegie Institution for Science, 5241 Broad Branch Road, NW, Washington, DC 20015, USA}
\email{nwallack@carnegiescience.edu}

\author[0000-0002-9479-2744]{Mykhaylo Plotnykov}
\affiliation{Department of Astronomy \& Astrophysics, University of Toronto; Toronto, Ontario, Canada M5S 3H4}
\email{mykhaylo.plotnykov@mail.utoronto.ca}

\author[0009-0000-8327-2631]{Alex McGinty}
\affiliation{Atmospheric, Oceanic, and Planetary Physics, Department of Physics; University of Oxford, Oxford OX1 3PU, UK}
\email{alex.mcginty@physics.ox.ac.uk}

\author[0009-0003-5977-9581]{Samuel Boucher}
\affiliation{Département de physique and Institut Trottier de recherche sur les exoplanètes, Université de Montréal, C.P. 6128, Succ. Centre-ville,
Montréal, H3C 3J7, Québec, Canada}
\email{samuel.boucher.3@umontreal.ca}

\author[0009-0009-6098-296X]{Bo Peng}
\affiliation{Department of Earth and Planetary Sciences, Stanford University; Stanford, CA, 94305, USA}
\affiliation{Department of Astronomy \& Astrophysics, University of Toronto; Toronto, Ontario, Canada M5S 3H4}
\email{bpengeps@stanford.edu}

\author[0000-0003-3993-4030]{Diana Valencia}
\affiliation{Department of Astronomy \& Astrophysics, University of Toronto; Toronto, Ontario, Canada M5S 3H4}
\email{diana.valencia@utoronto.ca}

\correspondingauthor{e.n.postolec@rug.nl}

\begin{abstract}

Rocky planets evolve through the exchange of volatiles between their interiors and atmospheres, an interplay still poorly constrained by observations. Remarkably, highly irradiated ultrashort-period (USP) exoplanets may offer a window into this exchange -- some retain low bulk densities compatible with volatile-rich envelopes surrounding rocky interiors, indicating possible secondary atmospheres. TOI-561\,b is a prime example, with a bulk density of $4.3\pm0.4$\,g\,cm$^{-3}$ and recent JWST observations favoring a thick volatile atmosphere overlying a dayside magma ocean. 
Here, we investigate the evolutionary pathways allowing TOI-561\,b to retain a substantial atmosphere over gigayears using the \texttt{PROTEUS} coupled interior--atmosphere framework. We explore different core radius fractions, Bond albedos, atmospheric escape efficiencies, mantle redox states, and initial C--H--O--N--S volatile inventories, under in situ evolution and late inward migration. 
Over half of our simulations leave a bare interior too dense to match observations. Successful cases favor a volatile-rich origin ($\lesssim200$\,Earth oceans of hydrogen, S/H\,$\le10$, and N/H\,$\le1$), an oxidized mantle ($f$O$_2 \gtrsim \mathrm{IW}+4$), a small iron core ($\le 0.40$ for the core radius fraction), and low escape efficiency ($\epsilon \lesssim 10^{-3}$) in the hydrodynamic escape regime. At present, TOI-561\,b is consistent with a global magma ocean beneath a thick (surface pressure $\approx 10^{3}$--$10^{4}$\,bar), high mean molecular weight atmosphere ($38$--$60$\,g\,mol$^{-1}$). Two archetypes emerge, differentiated by bulk sulfur content: a CO$_2$-dominated and an SO$_2$-dominated atmosphere. Migration is viable but not required to reproduce the observations. Our study illustrates how interior--atmosphere coupling governs atmospheric retention on irradiated rocky planets.

\end{abstract}


\section{Introduction}

A central question in exoplanet science is whether rocky planets can sustain atmospheres over geological timescales \citep{Kreidberg2025PNAS,Lichtenberg2025Sci}. Theoretical studies suggest that rocky planets can retain secondary atmospheres through a balance between volatile outgassing from their interiors and atmospheric loss processes operating over geological timescales \citep{Kite-2021, Krissansen-Totton-2024, Maurice_2024, Lichtenberg_2025, Nicholls_2025_tidal_l9869, Nicholls_2025_Nature, Nicholls_2025_MNRAS, Meier2026MNRAS}. 
Early observations with the \textit{Hubble Space Telescope} and the \textit{Spitzer Space Telescope} provided tentative hints of atmospheres for some nearby rocky and super-Earth exoplanets, while placing important constraints on others, including 55\,Cancri\,e \citep{Demory_2016_mnras,Demory_2016_nature,Tsiaras_2016,Mercier_2022}, GJ\,1132\,b \citep{Southworth_2017,Swain_2021}, and the TRAPPIST-1 system \citep{de_Wit_2018}.
Most recently, the larger aperture and extended wavelength coverage of JWST have provided tentative evidence for atmospheres on a growing sample of rocky exoplanets, including 55\,Cancri\,e \citep{Hu_2024,Patel-2024,snellen2026}, LHS\,1478\,b \citep{August_2025}, TOI-431\,b \citep{Monaghan_2025}, LTT\,1445A\,b \citep{Wachiraphan_2025}, HD\,3167\,b \citep{Coy_2026}, and the unusually underdense ultrashort-period planet TOI-561\,b \citep{teske_2025}. Escaping helium has also been detected from the super-Earth LHS\,1140\,b using ground-based high-resolution transit spectroscopy \citep{Cherubim_2026_science}. These findings motivate investigations into the processes that govern the origin, evolution, and persistence of atmospheres on irradiated rocky worlds.
    
Among the most challenging environments for atmospheric survival are ultrashort-period (USP) exoplanets, small rocky planets (of radius\,$R_{\rm p} \le 2 R_\oplus$) orbiting their host stars with periods shorter than a day. Exposed to intense stellar irradiation, many USP planets are expected to host surface magma oceans \citep{boukare_2022,Meier_2024}. Recent studies suggest that lava planets may have undergone inward migration \citep{Ogihara_2015, Petrovich_2019, Pu_2019, Lee_chiang_2017, Schmidt_2024, brandenberger_2026}, enabling them to store volatiles within their interior before reaching their present highly irradiated orbits. Subsequent outgassing could then replenish or sustain secondary atmospheres over long timescales \citep{Lichtenberg_2025,Lichtenberg2025Sci}. USP planets are therefore promising targets for the detection of secondary atmospheres, despite escape theory predicting that their atmosphere would be stripped away in the early planet evolution stage \citep{Zahnle_catling_2017_cosmic_shoreline,Owen_2019}.   
    
TOI-561\,b is a low-density USP \citep[period\,$P \sim 0.44$\,days,][]{Lacedelli_2022,Brinkman_2023,Patel_2023,Piotto_2024}. Its low dayside emission suggests a thick volatile atmosphere \citep{teske_2025}, and its 3--5\,$\mu$m phase curve indicates a high albedo as heat redistribution alone cannot explain the observations \citep{Boucher_2026}. The inferred bulk density $\rho_{\mathrm{obs}} =  4.3049^{+0.4411}_{-0.4216}$\,g\,cm$^{-3}$  \citep{Brinkman_2023, Patel_2023} is inconsistent with a purely rocky composition and suggests the presence of a significant volatile component \citep{Peng_Valencia_2024}. TOI-561\,b orbits around an old \citep[$11^{+2.8}_{-3.5}$\,Gyr,][]{Lacedelli_2022} G-type star in the thick Galactic disk. The host star is iron-poor \citep[Fe/H=$-0.40\pm0.05$\,dex, ][]{Lacedelli_2022} and enhanced in $\alpha$-elements \citep[e.g. O, Mg, Si; $\alpha$/H=$0.23\pm0.05$\,dex, ][]{weiss2021}, indicating a chemical composition distinct from that of the Solar System. Differing abundances of refractory and volatile elements can influence both the initial composition of the protoplanetary disk and the subsequent formation pathway of the planet \citep{Putirka_2019, Krijt2023ASPC, Spaargaren_2025}. TOI-561\,b provides a useful test  for understanding the origin and long-term evolution of volatile-rich USP planets.
    
In this work, we investigate which evolutionary pathways govern long-term atmospheric stability on the highly irradiated rocky exoplanet TOI-561\,b. We present the coupled interior--atmosphere framework used to model possible evolutionary pathway for TOI-561\,b in Sect.\,\ref{section methods}. Several possible evolutionary scenarios are explored in Sect. \ref{section results}. In Sect.\,\ref{section discussion}, we discuss the implications of our results in the context of observational constraints, as well as their relevance for the formation history of TOI-561\,b. We present our main conclusions in Sect.\,\ref{section conclusion}.
    
\section{Methods}
\label{section methods}

We aim to explore evolutionary pathways for TOI-561\,b that reproduce its present-day bulk properties to first order while remaining consistent with a present-day atmosphere, as suggested by JWST observations \citep{teske_2025}. 

Our simulations model the epoch of the secondary atmosphere of TOI-561\,b, set by outgassing and escape, and assume that any primordial H/He envelope was lost beforehand. We therefore do not explicitly explore an evaporated-core origin in the classical sense that is typically seen in the exoplanet astronomy literature, in which the planet began significantly more massive and lost a substantial primordial envelope to reach its present mass. In \texttt{PROTEUS}, the chemistry and envelope composition are set via the oxidation state and the total H--C--N--S abundances. Therefore, a higher volatile abundance and lower oxidation state can be interpreted as modeling a gas-dwarf scenario, while a higher oxidation state would fall in line with a water-world scenario. However, our maximal volatile abundances would need to be increased to reach these end-member scenarios and we thus cannot constrain in detail the maximally-possible volatile abundances.

Within this scope, we explore a broad parameter space of planetary properties and initial volatile inventories for  two evolutionary scenarios: in-situ evolution at the present-day orbit (scenario A) and late inward migration from a wider orbit (scenario B).

\subsection{Coupled interior--atmosphere modeling framework}
\label{section methods proteus}

We use the one-dimensional (1D) coupled interior–atmosphere \texttt{PROTEUS}\footnote{\url{https://proteus-framework.org}} framework \citep{Lichtenberg_2021_JGRP, Nicholls_2024_JGRP, Nicholls_2025_tidal_l9869, Nicholls_2025_MNRAS,Calder2026MNRAS,Sastre2026arXiv,vanDijk2026PSJ} to simulate the evolution of the USP TOI-561\,b, up to present-day ($\sim10^{10}$\,yr).
\texttt{PROTEUS} is a modular simulation and inference \citep{nicholls_constrainin_2026} framework that couples individual modules self-consistently to simulate planetary evolution, accounting for mantle dynamics, volatile exchange, atmospheric structure, atmospheric escape, stellar irradiation, and atmospheric chemistry. We set the initial volatile content for C--H--N--S through their abundances directly and for O through oxygen fugacity $f$O$_2$ imposed on the whole system, which sets the atmospheric and dissolved oxygen contents through equilibrium chemistry.  Bulk volatile inventory decreases over time through hydrodynamic escape to space, while volatiles are continuously exchanged between the mantle and the atmosphere. We model the magma-ocean era and do not treat the solid-state dynamics of a fully crystallized mantle. This does not affect our conclusions, as all observationally consistent cases remain global magma oceans at present day.

We resolve the 1D dynamic interior evolution using the \texttt{SPIDER} module \citep{Bower_2018, Bower_2019, Bower_2022}. Following \citet{Salvador_2023}, \citet{Nicholls_2024_JGRP, Nicholls_2025_Nature}, and \citet{Carone_2025}, we begin each simulation from a global magma ocean, as expected for a rocky planet following accretion. This assumption translates to an initial specific entropy of $S_{\mathrm{init}} = 4000\,\mathrm{J\,K^{-1}\,kg^{-1}}$, above the silicate liquidus throughout the mantle so that the global melt fraction, $\phi_{\rm global} = 1$.
\texttt{SPIDER} evolves depth-dependent profiles of specific entropy, temperature, melt fraction and composition during the magma ocean era. It accounts for heat transport (convection), radiogenic heating, and for phase mixing, separation, and gravitational settling of solid and melt. Core cooling is imposed by a lower boundary heat flux; the core itself is not
explicitly modeled, but treated as an inert, volatile-free reservoir. 
The mantle is treated as a single silicate component (MgSiO$_3$) using the equation of state of \citet{WOLF2018}. We adopt melting curves that give bottom-up crystallization: solidification proceeds from the base of the mantle upward \citep{Bower_2019,Nicholls_2024_JGRP}.
Given the short orbital period of TOI-561\,b \citep[$P = 0.44$\,days;][]{Patel_2023,Farhat2025ApJ}, tidal dissipation may contribute to the planet's internal energy budget and overall evolution \citep{Herath_2024_mnras,Nicholls_2025_tidal_l9869, vanDijk2026PSJ}. Tidal heating is not considered in this work and would only further support our conclusions, as discussed in Sect.\,\ref{section discussion physical atmosphere}.

To model volatile exchange between the interior and the atmosphere, we use the \texttt{CALLIOPE} module \citep{Bower_2022, SOSSI_2023, Nicholls_2024_JGRP, Nicholls_2025_MNRAS}. Volatile partitioning between the magma interior and the atmosphere is calculated assuming equilibrium chemistry and experimentally derived solubility laws \citep{DIXON_1995, Chase1998AIP, ONEILL_2002, ARDIA_2013, ARMSTRONG2015, Dasgupta_2022, Gaillard_2022} for major volatile species, including H$_2$O, CO$_2$, N$_2$, S$_2$, SO$_2$, H$_2$S, NH$_3$, H$_2$, CH$_4$, CO, and O$_2$. 
At each time-step, \texttt{CALLIOPE} recomputes the volatile exchange from the melt fraction and surface temperature supplied by \texttt{SPIDER}, updating the surface pressure $P_{\rm surf}$ \citep{Bower_2022, Gaillard_2022}. The redox speciation is governed by the prescribed mantle oxygen fugacity $f$O$_2$ \citep{Ortenzi_2020,Gaillard_2022,SOSSI_2023}, parametrized as an offset $\Delta$IW relative to the temperature-dependent iron–w\"ustite (IW) buffer reaction \citep{ONEILL_2002}.
A thin conductive boundary layer ($1\,\mathrm{cm}$ and thermal conductivity of $2.0\,\mathrm{W\,m^{-1}\,K^{-1}}$) regulates mantle-to-surface heat transport and modulates volatile exchange at the magma ocean--atmosphere interface \citep{ElkinsTanton2008,Hamano-2015}.

The atmospheric structure is updated at each \texttt{PROTEUS} time step, using the \texttt{AGNI} module \citep{Nicholls_agni_2025, Nicholls_2025_MNRAS}, which solves for energy balance in a 1D atmosphere, allowing for convective and radiative layers. The top of the atmosphere is set at $10^{-6}$\,bar and discretized using a 50-level vertical grid. The lower boundary is set at the outgassed surface pressure $P_{\rm surf}$ computed by \texttt{CALLIOPE}. Volatile condensation is included as rainout, while radiative effects from clouds, aerosols, and rock vapor are not considered in this study despite their potential relevance for the evolution of TOI-561\,b \citep{Piette_2023,Janssen_2026}. We address this limitation in Sect.\,\ref{section model limitations}.
Radiative fluxes are computed with the \texttt{SOCRATES} code \citep{edwards_studies_1996, amundsen_radiation_2014, Sergeev_2023, Manners_2024}, which \texttt{AGNI} uses to solve the atmospheric $T$($P$) structure. 
Fluxes in each layer are evaluated over 48 spectral bands using the Honeyside spectral file \citep{Nicholls_2025_honeyside48}: $k$-coefficients fitted to opacity cross-sections from the DACE database \citep{Grimm-2021}, which incorporates ExoMol \citep{Tennyson_2016}, HITRAN \citep{GORDON2022,Gordon_2026}, and HITEMP \citep{Rothman_2010} linelist data. Line absorbers are H$_2$O, H$_2$, CO$_2$, CO, CH$_4$, N$_2$, NH$_3$, SO$_2$, N$_2$O, O$_3$, HCN, and H$_2$S, supplemented by nine collision-induced continua. For further details on the opacity
treatment, see \cite{Nicholls_2024_JGRP} and \cite{Nicholls_2025_MNRAS}. We do not consider Rayleigh scattering but instead investigate several prescribed Bond albedos.

Atmospheric escape is computed with an energy-limited (EL) formalism \citep{Watson_1981, Erkaev_2007, lopez-fortney-2013} through the \texttt{ZEPHYRUS} module \citep{Postolec_2026}. In our model, the mass-loss rate scales as $\dot{M}_{\rm EL}\propto  \epsilon R_\mathrm{XUV}(t)^3 F_\mathrm{XUV}(t) /M_\mathrm{p}$, where $\epsilon$ is the heating efficiency and $R_\mathrm{XUV}$ the X-ray and extreme ultraviolet (XUV) absorption radius. The escape is driven by the stellar XUV flux $F_{\rm XUV}(t)$, provided by the \texttt{MORS} module from stellar evolution tracks \citep{Johnstone_2021}, and occurs at $R_{\rm XUV}(t)$, the radius at which the atmosphere becomes optically thick to XUV radiation, computed by \texttt{AGNI}.
Given its intense irradiation ($\sim3900$ times the present-day insolation of Earth), TOI-561\,b is expected to undergo hydrodynamic escape, in which the outflow is non-fractionating and species are lost together. At each time step, the mass-loss rate is partitioned among the atomic species in proportion to their outgassed abundance, updating the planetary bulk inventory.  This treatment assumes that molecules are photodissociated into atoms in the upper atmosphere.
We explore escape efficiencies from $\epsilon = 10^{-3}$ to $0.1$, below the value commonly adopted for super-Earths \citep[$\epsilon \approx 0.15$;][]{Luger_Barnes_2015, Kubyshkina-2018, Owen_2019, Cherubim_2024}. Such low efficiencies are appropriate for the modeled high mean molecular weight (MMW) atmospheres of TOI-561\,b, whose small scale heights reduce $R_{\rm XUV}$ and hence the mass-loss rate (Sect.\,\ref{section discussion physical atmosphere}).

The vertical chemical structure of the atmosphere is computed as a post-processing step, applied to the atmosphere at the final, converged time step. We use the \texttt{VULCAN} module \citep{Tsai_2021, Tsai_2023} to compute pressure-dependent chemical abundances using a C--H--O--N--S photochemical kinetics network. The outgassed composition provides the initial condition of the kinetic model, and zero-flux boundary conditions are imposed at the top and bottom of the atmosphere. \texttt{VULCAN} includes disequilibrium chemistry through vertical mixing, molecular diffusion, and photodissociation. The eddy diffusion profile $K_{\rm zz}$ is taken directly from the convective mixing computed by \texttt{AGNI}, reaching ${\sim}10^{9}$--$10^{11}\,\mathrm{cm^{2}\,s^{-1}}$. Condensation and settling are not included. These processes contrast with the vertically well-mixed composition assumed during the evolution, where each species volume mixing ratio (VMR) is held constant with altitude, although the bulk composition itself evolves continuously through outgassing and escape.
The evolutionary energy balance in this work assumes an isochemical atmosphere set by the outgassed composition; this assumption is discussed further in Sect.\,\ref{section model limitations}.

Stellar parameters for TOI-561 (effective temperature, surface gravity, metallicity, and $\alpha$-element abundance) are taken from \citet{Lacedelli_2022} \citep[consistent with Table A.1 of][]{teske_2025} and used to select the corresponding high-resolution PHOENIX spectrum \citep{Husser_2013_phoenix}. These parameters initialize the \texttt{MORS} module \citep{Johnstone_2021}, which interpolates the stellar evolution tracks from \cite{2013_spada} to compute the XUV irradiation history of TOI-561.
We assume a median (50th percentile) stellar rotation rate appropriate for an old solar-type star. We assume an initial stellar age of 0.1\,Gyr and then evolve the stellar spectrum with a temporal resolution of 1\,Myr, while the incident stellar flux is updated every 100\,yr.

We initialize the planetary mass at the observed value of $2.24\,M_{\oplus}$ \citep{Brinkman_2023} and evolve it through volatile replenishment from the interior and atmospheric escape. For the cases that reproduce the observations, the total mass changes by less than ${\sim}1\%$, far within the $\pm0.20\,M_{\oplus}$ (${\sim}9\%$) measurement uncertainty \citep{Brinkman_2023}. The planetary radius is not prescribed as an initial condition, but instead computed self-consistently as part of the model evolution. 
The observed density \citep[$\rho_{\mathrm{obs}} = 4.3049^{+0.4411}_{-0.4216}$\,g\,cm$^{-3}$;][]{Patel_2023}  independently supports our assumption that no primordial H/He envelope survives to the present day.

\subsection{In-situ vs late inward migration scenarios}
\label{section scenario a and b in methods}

We investigate two sets of evolutionary scenarios that could explain the present-day properties of TOI-561\,b. 

In the in-situ scenario A, the planet remains at its present-day orbital separation throughout its entire evolution following its formation in the protoplanetary disk and a potential giant impact phase.
Scenario A remains consistent with substantial early inward migration, either disk-driven (type I or type II) during the disk lifetime \citep{Ward_1997, Kley_2012, Baruteau_2014, brandenberger_2026} or through orbital rearrangement during the giant-impact phase \citep[$\sim10^{1}$\,Myr;][]{Raymond_2018,Izidoro_2021}. We emphasize that in-situ here denotes the absence of late migration only, while scenario B explicitly models late migration occurring after disk dispersal.

In the late inward migration scenario B, we hypothesize that the planet forms at a larger orbital distance and migrates inward at a later stage of its evolution, consistent with proposed late-migration pathways for USP planets \citep{Lee_chiang_2017, Petrovich_2019, Schmidt_2024}. 
This narrative has the benefit of shielding the planet from intense escape driven by the star's infancy, migrating after the energetic saturation phase \citep[e.g.,][]{Jackson-2012} and potentially helping atmospheric survival \citep{Attia_2021,Attia_2025}. One plausible mechanism is high-eccentricity migration, in which the orbital eccentricity is pumped to high values, triggering strong tidal dissipation at periapsis, which leads to gradual circularization and inward migration over Gyr timescales \citep[e.g.,][]{Fabrycky_2009, Nagasawa_2011, Petrovich_2019}. 
Such eccentricity excitation can be driven by secular \citep[e.g.,][]{Hamers_2017} or resonant \citep[e.g.,][]{Chiang_2002} interactions with outer companions. TOI-561\,b belongs to a system of four confirmed planets, whose outer planets d and e remain in a $3{:}1$ mean-motion resonance \citep{Lacedelli_2022}, indicative that the planets may have migrated inward together as a resonant chain \citep[e.g.,][]{Izidoro_2017, Mills_2016}. Such chains can become unstable and break from the inside out \citep{Izidoro_2021}, driving the innermost planet inward through one or more transient high-eccentricity episodes. The persistence of the resonance between planet d and e today suggests this may still be ongoing, i.e. that the migration of TOI-561\,b occurred late during its evolution. We treat its late inward displacement as a prescribed change in orbital distance, rather than modeling the migration dynamics explicitly (which is beyond the scope of this work).

For the late inward migration scenario B, we consider a pathway in which TOI-561\,b undergoes a high-eccentricity phase during its evolution, resulting in a late inward orbital migration. The eccentricity $e(t)$ is assumed to follow the circularization model of \citet[][their Eq.\,(8)]{Correia_2020}, which can be seen as the simplest physically motivated eccentricity-damping formulation:
\begin{equation}
    e(t) = e_0 \exp\left(-\frac{t - t_0}{\tau}\right)
    \label{equation eccentricity evolution}
\end{equation}
where $e_0$ is the initial eccentricity at the onset migration time $t_0$, and $\tau$ is the characteristic migration timescale. Assuming conservation of orbital angular momentum $h$ during the tidal circularization track, as in \cite{Socrates_2012}, we approximate
\begin{equation}
    h^2 \propto a(t)\left(1 - e(t)^2\right)
    \label{equation angular momentum conservation}
\end{equation}
where $a(t)$ is the semimajor axis time evolution, to be a constant quantity. Since the orbit of TOI-561\,b is circularized at the present epoch \citep{Lacedelli_2022,Brinkman_2023,Patel_2023,Piotto_2024}, we have $e_{\rm f} \approx 0$, $a \rightarrow a_{\rm f}$, where $e_{\rm f}$ is the final eccentricity and $a_{\rm f}$ is the final semi-major axis corresponding to the present-day orbit of TOI-561\,b, i.e. $a_{\rm f}=0.0106$\,au \citep{Patel_2023}. We have $a(t)\left(1 - e(t)^2\right) = a_{\rm f}$, which yields
\begin{equation}
    a(t) = \frac{a_{\rm f}}{1 - e_0^2 \exp\left(-\frac{2 (t - t_0)}{\tau}\right)}
    \label{equation sma evolution}
\end{equation} 

We consider initial semi-major axes $a_0=0.029$\,au and $a_0=0.0889$\,au, corresponding to initial orbital periods $P_0 \approx 2$\,days and  $P_0 \approx 10.8$\,days, from which the planet migrates inward to its present-day period $P \sim 0.44$\,days \citep{Patel_2023}. 
The two initial periods we adopt span the low-order mean-motion resonances with the inner companion c, with the larger value coinciding with c's present-day orbit ($P_c = 10.8$\,days), consistent with an origin in a resonant chain. Under this assumption, the initial eccentricity corresponding to the adopted initial and final orbital separations is given by
\begin{equation}
e_0 = \sqrt{1 - \frac{a_{\rm f}}{a_0}}.
\label{equation injected eccentricity}
\end{equation} 
Prior to the onset of migration at $t_0$, the planet is held on a circular orbit ($e=0$) at its initial semi-major axis $a_0$. At $t_0$, a high-eccentricity event (e.g., close encounter, resonance crossing) excites the orbit to $e_0$, after which tidal dissipation circularizes it ($e\rightarrow e_{\rm f}\approx0$) while the semi-major axis contracts toward $a_{\rm f}$, as shown in Appendix Fig.\,\ref{plot appendix sma ecc evolution}.

\subsection{Parameter-space exploration}

\begin{deluxetable}{lcc}
\tablecaption{Explored parameter space for the two evolutionary scenarios investigated for
TOI-561\,b.\label{table parameter space exploration}}
\tablewidth{0pt}
\tablehead{
\colhead{Parameter [unit]} & \colhead{Symbol} & \colhead{Values}
}
\startdata
\cutinhead{Scenario A: in situ}
Core-radius fraction [non-dim.]               & CRF           & 0.30, 0.35, 0.40, 0.55 \\
Bond albedo [non-dim.]                        & $A_{\rm b}$         & 0.0, 0.6, 0.9 \\
Escape efficiency factor [non-dim.]           & $\epsilon$    & $10^{-3}$, $10^{-2}$, $10^{-1}$ \\
Oxidation state [$\Delta$IW]\tablenotemark{a}     & $f$O$_2$      & $+0$, $+2$, $+4$ \\
Total H inventory [Earth oceans]\tablenotemark{b} & [H]           & 100, 200, 400, 1000 \\
C/H ratio [MMR]\tablenotemark{c}                  & C/H           & 0.1, 1.0, 2.0, 10 \\
N/H ratio [MMR]\tablenotemark{c}                  & N/H           & 0.015, 0.1, 1.0, 10 \\
S/H ratio [MMR]\tablenotemark{c}                  & S/H           & 0.1, 1.0, 10, 100\\
\cutinhead{Scenario B: late inward migration}
Initial semimajor axis [au]                       & $a_0$         & 0.0290, 0.0889 \\
Migration timescale [yr]                          & $\tau$        & $10^{6}$, $10^{8}$, $10^{10}$ \\
Migration onset time [yr]                         & $t_0$ & $10^{8}$, $10^{9}$, $10^{10}$ \\
Escape efficiency factor [non-dim.]          & $\epsilon$    & $10^{-3}$, $10^{-2}$, $10^{-1}$ \\
C/H ratio [MMR]\tablenotemark{c}                  & C/H           & 1.0, 10 \\
S/H ratio [MMR]\tablenotemark{c}                  & S/H           & 0.1, 10 \\
\enddata
\tablenotetext{a}{Log$_{10}$ of the oxygen fugacity relative to the iron--w\"ustite (IW) buffer \citep{ONEILL_2002}.}
\tablenotetext{b}{1 Earth ocean $= 1.39\times10^{21}$~kg of H$_2$O \citep{Genda_2008}.}
\tablenotetext{c}{Mass mixing ratio in the mantle/atmosphere system.}
\tablecomments{Scenario B additionally adopts the interior and volatile
parameters listed under Scenario A.}
\end{deluxetable}

With our framework and evolution scenarios set, we explore a broad parameter space to identify conditions that are consistent with the observed properties of TOI-561\,b. 

For the in-situ scenario A, the varied input parameters (summarized in Table\,\ref{table parameter space exploration}) are the core radius fraction (CRF), Bond albedo $A_{\rm b}$, atmospheric escape efficiency $\epsilon$, relative oxygen fugacity $f$O$_2$, and initial volatile inventory including hydrogen content in Earth oceans \citep[EO;][]{Genda_2008}, C/H, N/H, and S/H mass ratios. 
In \texttt{PROTEUS}, the planet consists of a silicate mantle overlying an inert metallic iron core, treated as an inner boundary condition rather than explicitly modeled. We consider core-radius fractions lower than that of Earth \citep[CRF=0.55;][]{Dziewonski-1981}, motivated by the iron-poor composition and the enhancement of alpha-elements (Mg, Si, O) in the host star. TOI-561 has sub-solar refractory abundance ratios \citep[e.g. Fe/Mg\,$\approx0.52$ and Fe/Si\,$\approx0.60$;][]{weiss2021}, expected to correspond to a lower bulk core mass than Earth, and hence a smaller core radius fraction \citep{Adibekyan_2021, Adibekyan_2024, Brinkman_2024,Plotnykov_2026}.
We explore a range of Bond albedos of $A_{\rm b} = 0$, $0.6$, and $0.9$. The dark case ($A_{\rm b}=0$) represents a planet with a clear-sky and magma-ocean surface with low-reflectivity \citep{Essack-2020}. The higher values are indicative of reflective cloudy atmospheres, motivated by the high albedo inferred for TOI-561\,b \citep[$A_{\rm b}\sim0.6$;][]{Boucher_2026}. However, cloud formation processes are not explicitly modeled here.
We also explore a wide range of values of the escape efficiency parameter ($\epsilon$) within the hydrodynamic escape regime from $10^{-3}$ to $10^{-1}$. The motivation and implications of this choice are discussed in Sect. \ref{section discussion physical atmosphere}. We explore volatile-rich inventories with initial hydrogen content ranging from $100$--$1000$\,EO, corresponding to bulk water mass fractions of $\approx1$--$10\,$wt\% for TOI-561\,b, one to two orders of magnitude larger
than that of the Earth, whose bulk water content is only a few oceans \citep[$\lesssim0.1$\,wt\%;][]{Marty_2012}. The sampled elemental ratios bracket the bulk silicate Earth (BSE) values, which by mass are C/H\,$\approx1.3$, N/H\,$\approx0.03$, and S/H\,$\approx2.7$ \citep{Hirschmann_2016}. Our C/H grid ($0.1$--$10$) and S/H grid ($0.1$--$100$) thus span roughly $0.1$--$8\times$ and $0.04$--$40\times$ terrestrial, centered near the BSE value, while N/H ($0.015$--$10$) ranges from approximately terrestrial to strongly nitrogen-enriched. Oxidized initial conditions \citep[$f$O$_2 \geq \rm IW$, up to the terrestrial
upper-mantle value of IW$+4$;][]{Frost_McCammon_2008} and enhanced primordial volatile inventories are also explored to assess a wide range of plausible interior–atmosphere formation and evolution histories. 

The values in Table\,\ref{table parameter space exploration} are sampled across several sub-grids rather than as a single full grid, so not a full Cartesian parameter space of all possible input parameters are explored jointly. This reflects our aim of a sensitivity study mapping how the model responds to each parameter. Enumerating every combination is computationally expensive (typically several hours, up to $\sim5$\,days per simulation) and unnecessary to identify the dominant controls on the planet's evolution.

For the late inward migration scenario B, we vary the initial semimajor axis $a_0$ between the separation corresponding to a 2-day orbit ($a_0 = 0.0290$\,au) and the present-day orbit of planet TOI-561\,c ($a_0 = 0.0889$\,au at $P_0 = 10.8$\,days). We adopt planet c orbit as the upper bound of the tested semimajor axis because, in the TOI-561 system, planet c is the closest potential secular perturber of planet b that could excite its eccentricity and drive its inward migration. The lower value ($a_0=0.0290$\,au, $P_0\approx2$\,days) represents a moderately migrated short-period orbit interior to companion c, chosen as a representative lower bound on the pre-migration separation rather than a specific resonance.
The migration timescale $\tau$ sets the rate of circularization and is sampled from $10^{6}$ to $10^{10}$\,yr. In this parameterization, the injected eccentricity is fixed by the initial and final separations given by Eq.\,\ref{equation injected eccentricity}, where $e_0\approx0.80$ for $a_0=0.0290$\,au and $e_0\approx0.94$ for $a_0=0.0889$\,au. We discuss the implications of high-eccentricity events in Sect.\,\ref{section model limitations}. Thus $\tau$ controls only how rapidly the orbit circularizes, not the eccentricity magnitude. We test late migration onset time ($t_0=10^{8}$--$10^{10}$\,yr) compared to TOI-561\,b age \citep[$11^{+2.8}_{-3.5}$\,Gyr;][]{Lacedelli_2022}. This favors atmospheric retention by keeping the planet at lower instellation, and thus lower atmospheric escape rates.
The remaining interior and volatile parameters are held fixed at a fiducial configuration, chosen as representative of the volatile-rich, oxidized compositions we explore in scenario A: the core-radius fraction to 0.35, the Bond albedo to $A_b = 0.0$, the oxygen fugacity to $f$O$_2 = \rm IW +4$ and an initial volatile inventory of [H]\,$= 100$\,EO and N/H\,$= 0.1$ by mass. Fixing these parameters isolates the effect of the migration history on the final state of the planet. However, we note that migration could additionally enable atmospheric retention for interior and volatile configurations that fail in situ (by shielding the planet from intense escape rates), and thus opening regions of parameter space inaccessible in scenario A. We defer a joint exploration of migration with a broader interior--volatile parameter space to future work.

For the results presented here, we retain only simulations that run to convergence, i.e. that reach the adopted stopping time at system age \citep[$\sim$\,11\,Gyr;][]{Lacedelli_2022} with all coupled modules numerically stable throughout. This represents 78\% of the full parameter grid (1030 of 1323 simulations). The remaining runs terminate earlier due to numerical instability in the coupled interior--atmosphere system and are therefore excluded from the analysis. Non-convergence is spread evenly across the grid, aside from a modest excess at the most extreme volatile inventories (S/H\,$=100$, [H]\,$=1000$\,Earth oceans). As these extreme cases lie outside the regime that reproduces the observed bulk properties, their exclusion does not bias our results.

Bulk density estimates for TOI-561\,b range from $3.8\pm0.5$\,g\,cm$^{-3}$ \citep{Lacedelli_2022} up to $4.8\pm0.5$\,g\,cm$^{-3}$ \citep{Brinkman_2023}, all indicative of a volatile-rich planet with a rocky interior \citep{Zeng_2019}. We adopt the most precise values: bulk density $\rho_{\mathrm{obs}}\,=\,4.3049^{+0.4411}_{-0.4216}$\,g\,cm$^{-3}$ \citep{Patel_2023}, and radius $R_{\mathrm{obs}}\,=\,1.4195^{+0.0217}_{-0.0224}\,R_\oplus$ \citep{Patel_2023}, and mass $M_{\mathrm{obs}}\,=\,2.24 \pm 0.20 \,M_\oplus$ \citep{Brinkman_2023}.
The model radius ($R_{\rm mod}$) computed by \texttt{PROTEUS} is the transit radius, computed by \texttt{AGNI} at a fixed reference pressure of $20$\,mbar (representative of the level probed in transmission). The model bulk density is derived from the transit radius and the planetary mass. Both are therefore directly comparable to the transit-based observed quantities.
To evaluate consistency with observations, we apply a two-step criterion. We first require that a simulation retain an atmosphere at the end of the run (i.e., $P_{\mathrm{surf}} > 0$\,bar). Second, we require its bulk density, radius, and mass at the end of the run \citep[$11.0$--$11.5$\,Gyr, within the system age of $11^{+2.8}_{-3.5}$\,Gyr;][]{Lacedelli_2022} to match the observed values within their $1\sigma$ uncertainties.

\section{Results}
\label{section results}

Using our coupled interior--atmosphere framework, we identify the evolutionary pathways that allow TOI-561\,b to retain an atmosphere consistent with its observed present-day bulk properties.

\subsection{Atmosphere retention across parameter space}

\begin{figure*}
    \centering
    \includegraphics[width=\textwidth]{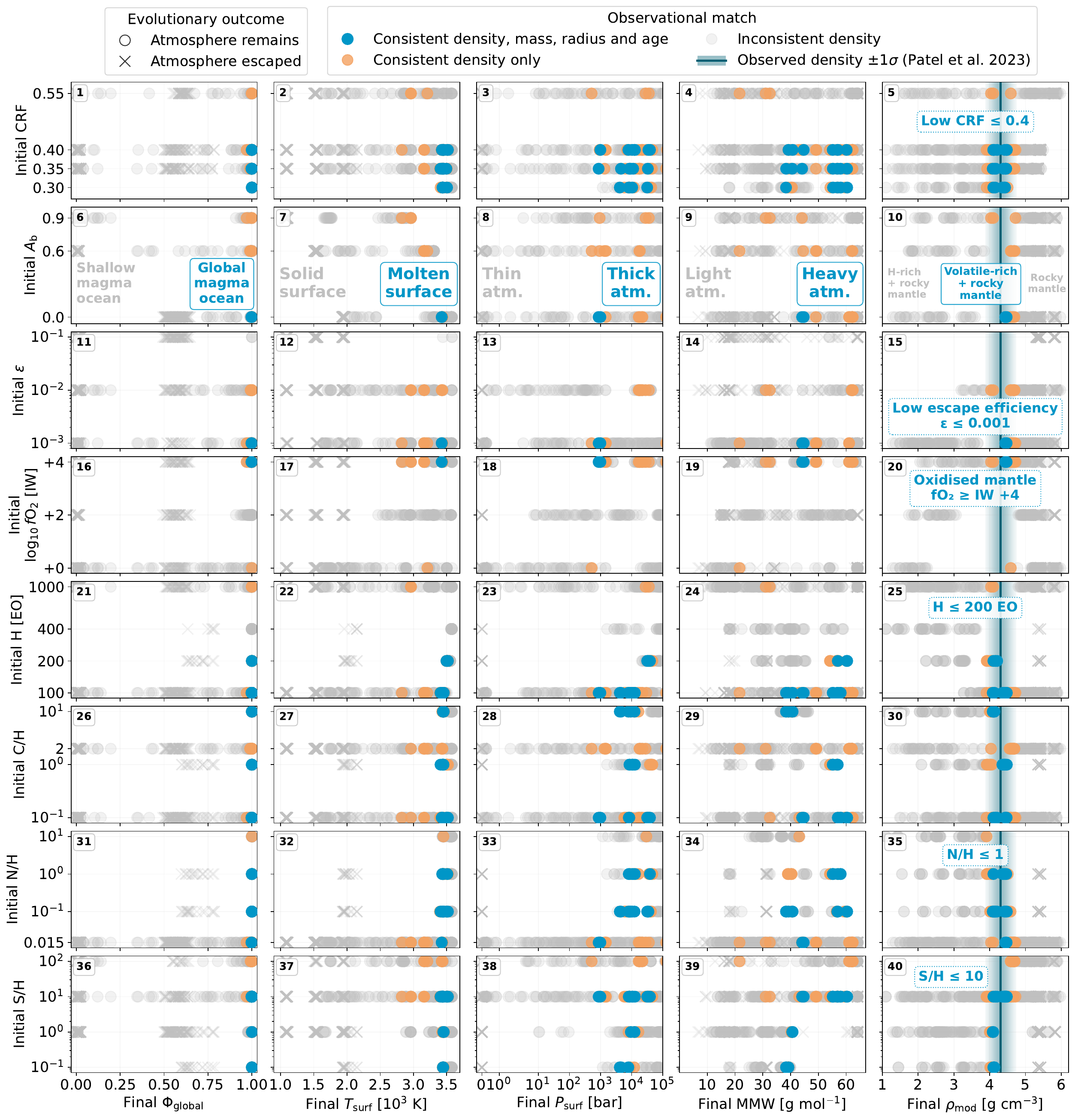}
    \caption{Final planetary properties (columns) as a function of the initial grid parameters (rows) for the in-situ scenario A. Properties are measured at the end of each run, at the system age \citep[$11.0$--$11.5$\,Gyr, within the observed $11^{+2.8}_{-3.5}$\,Gyr;][]{Lacedelli_2022}. Symbol shape marks the evolutionary outcome (circles: atmosphere retained; crosses: complete atmospheric loss). Color indicates agreement with observations (blue: consistent with observed density, mass, radius, and age; orange: consistent with density only; gray: inconsistent with observations). In the last column, the blue dark line and shaded band show the observed bulk density and its $1\sigma$ uncertainty, $\rho_{\mathrm{obs}} = 4.3049^{+0.4411}_{-0.4216}$\,g\,cm$^{-3}$ \citep{Patel_2023}. Blue-outlined boxes summarize the observationally consistent (blue) models. Dotted boxes in the last column give the inferred constraints on the initial planetary properties (favored parameter ranges), and solid boxes in the second row describe the corresponding present-day state. No box is drawn for C/H because successful models span its full sampled range, so it remains unconstrained. Gray labels denote the opposite, unsuccessful regime.}
    \label{figure grid scatter point}
\end{figure*}

Figure\,\ref{figure grid scatter point} shows the simulation-grid results for the in-situ scenario A of TOI-561\,b. We vary eight input parameters (scenario A, Table\,\ref{table parameter space exploration}): core-radius fraction (CRF), Bond albedo $A_b$, escape efficiency $\epsilon$, upper-mantle oxygen fugacity $\Delta$IW, bulk hydrogen inventory H in Earth oceans [EO], and the bulk C/H, N/H and S/H mass ratios. Each row shows how a final planetary property (columns), measured at the end of the run, responds to one input parameter. Blue circles indicate simulations matching current day observations (density, mass, radius and age) at the end of the simulation, while gray cross correspond to planet that lose their atmosphere before the current age ($\sim$\,11\,Gyr).

Of the 954 completed simulations for the in-situ scenario A, 499 ($\sim$52\%) lose their atmosphere entirely before the system age (crosses), while 455 retain an atmospheric envelope (circles). Only 19 ($\sim$2\%) simulations reproduce all four bulk observables simultaneously (blue), while 36 match only the observed density, falling outside of the mass or radius window (orange). This fraction reflects how tightly the joint bulk constraints select the input parameters, rather than a general probability of atmospheric retention for USP planets, since our sub-grids are deliberately sampled to favor retention. The remaining 400 retained cases are inconsistent with the density constraint (gray circles).
No atmosphere-stripped case matches any of the bulk constraints: complete loss leaves a bare interior with $\rho_{\rm mod} \approx 5.5$\,g\,cm$^{-3}$, far too dense to match the observed value.

Among the cases that retain an atmosphere but fail to match observational constraints (gray), two opposing regimes emerge. Planets that are too high in density host sulfur-rich, high MMW atmospheres (S/H\,$\ge10$, panel\,40 in Fig.\,\ref{figure grid scatter point}), dominated by SO$_2$ (MMW\,$\approx60$\,g\,mol$^{-1}$). Because such heavy atmospheres have a small scale height, their final radii are compact ($R_{\mathrm{mod}}\approx1.3\,R_\oplus$), lying below the observed radius $R_{\rm obs} =  1.4195^{+0.0217}_{-0.0224}\,R_\oplus$ \citep{Patel_2023}. Such planets end up too dense despite retaining a substantial volatile envelope. On the contrary, planets that are too low in density begin volatile-rich and retain a massive envelope, favored at low escape efficiency ($\epsilon=0.001$, panel\,15). Such atmospheres inflate the final radius ($R_{\mathrm{mod}}\approx1.6\,R_\oplus$) and decrease the bulk density below the observed range. 

Only planets that retain an atmosphere in our model match present-day observations of TOI-561\,b. TOI-561\,b therefore requires a substantial volatile inventory to survive $\sim$11\,Gyr of escape, but not so much that the planet remains over-inflated at the present day. 
Our planetary evolution models provide a physical basis and justification for the observational metrics that point to atmosphere retention on TOI-561\,b \citep{teske_2025,Boucher_2026}.

\subsubsection{Inferred initial planetary properties of TOI-561\,b}

For the in-situ scenario A, our results favor a core-radius fraction smaller than Earth's \citep[CRF $=0.55$;][]{Dziewonski-1981}. The successful cases occupy the low end of our sampled range (CRF\,$=0.30$--$0.40$; panels\,1--5 of Fig.\,\ref{figure grid scatter point}), while the Earth-like value is excluded. Thus, valid cases with CRF\,$\lesssim0.40$ for TOI-561\,b should be interpreted as an upper bound rather than a preferred value since our grid does not sample CRF\,$<0.30$. 
A smaller core than Earth for TOI-561\,b is consistent with the sub-solar refractory ratios of the host star (low Fe/Mg and Fe/Si), which combine its low iron content with its $\alpha$-element enhancement \citep{weiss2021,Lacedelli_2022}. It also implies a correspondingly larger silicate mantle fraction, leading to a lower bulk density. In our setup the volatile inventory is fixed in absolute units of Earth oceans for hydrogen (and consequently all the mass mixing ratios), independent of mantle mass. This treatment reflects the interior structure rather than a larger volatile budget. However, a larger silicate mantle could physically host a proportionally larger volatile reservoir \citep{Lichtenberg_2021_JGRP, Dorn_lichtenberg_2021, Bower_2022}, further favoring sustained outgassing and atmospheric retention. We do not capture this effect here, since our volatile budget does not scale with mantle mass.

Under our cloud-free assumption, only planets with an initial zero Bond albedo (Fig.\,\ref{figure grid scatter point}, panels\,6--10) reproduce the observations, consistent with the low reflectivity of magma-ocean surfaces \citep{Essack-2020}. This finding differs from previous studies \citep{teske_2025,Boucher_2026}, which find highly reflective atmospheres ($A_{\rm b} = 0.6$), potentially indicative of clouds. This discrepancy arises because we model only cloud-free, transparent atmospheres, not accounting for Rayleigh scattering, which yield low $A_{\rm b}$. 
Imposing a higher albedo ($A_{\rm b}=0.6$--$0.9$) instead reduces the absorbed instellation in our model, cooling the surface (median $T_{\rm surf}$ from ${\sim}3450$\,K at $A_{\rm b}=0$ down to ${\sim}2850$\,K at $A_{\rm b}=0.9$) and slightly shrinking the transit radius. None of these cases matches all observational constraints, only density (orange dots in panel\,7).
We therefore cannot place a strong constraint on the Bond albedo of TOI-561\,b from this work.

An oxidized mantle is favored for TOI-561\,b (Fig.\,\ref{figure grid scatter point}, panels\,16--20), with the successful cases at the most oxidized value we sampled ($f$O$_2$\,=\,IW+4). Since this is the ceiling of our grid, it represents a lower bound ($f$O$_2 \gtrsim \mathrm{IW}+4$) rather than a preferred value. More oxidized conditions are not explored here but remain possible.
This oxidation state may reflect the planet's formation history, for example through inward migration in the protoplanetary disk \citep{Lee_chiang_2017,Carrera_2019} and the accretion of oxygen-rich material during planetary accretion \citep{Doyle_2019,Ortenzi_2020}, such as water ice and outer disk volatiles \citep{Bitsch2019AA,Lichtenberg2019NatAs,Lichtenberg2022ApJL,Krijt2023ASPC,Krijt2025ApJL}. Alternatively, mantle oxidation could have been generated internally \citep{Schaefer_2024} through iron disproportionation reactions \citep{Frost_2004,Wade_2005,Armstrong_2019,Hirschmann_2022,nicholls2026chili}. These possible origins are discussed further in Sect.\,\ref{section discussion formation}. 

Retaining an atmosphere at the present-day orbital separation requires a very low escape efficiency, $\epsilon \sim 0.001$ (Fig.\,\ref{figure grid scatter point}, panels\,11--15), i.e. the lowest value we sampled. As for the core-radius fraction and oxygen fugacity, this favored value sits at a grid boundary and is best read as an upper limit ($\epsilon \lesssim 0.001$) rather than a located optimum. The plausibility of such a low value is discussed in Sect.\,\ref{section discussion physical atmosphere}. At higher escape efficiency ($\epsilon = 0.1$), almost all atmospheres are lost.

Sustaining an atmosphere over 11\,Gyr requires a high initial volatile inventory to balance intense atmospheric escape at short orbital period. An initial bulk hydrogen inventory of H\,$\lesssim 200$ Earth oceans (partitioned between the atmosphere and the mantle but not in the core) matches the present-day observations (Fig.\,\ref{figure grid scatter point}, panels\,21--25). Larger inventories (H\,$\ge400$\,EO) leave the planet over-inflated, with densities below the observed range ($\rho_{\rm mod}\approx1$--$3$\,g\,cm$^{-3}$, gray dots in panel\,25). Because our smallest sampled value (H\,$=100$\,EO) is also consistent, we can place an upper bound (H\,$\lesssim200$\,EO) on the initial volatile content.
Sulfur-rich compositions are preferred (S/H\,$\le10$) -- $14$ of the $19$ successful cases have S/H\,$=10$ -- though lower values are not excluded
(panels\,36--40).
The successful cases favor N/H\,$\le1$ (panels\,31--35). This spans from roughly terrestrial \citep[N/H\,$\approx0.12$;][]{Wang_2018} to ${\sim}8\times$ that value. The nitrogen-rich end of our grid (N/H\,$=10$) is disfavored. The initial C/H ratio is poorly constrained (panels\,26--30).

\subsubsection{Modeled present-day properties of TOI-561\,b}

The successful simulations all host a global, fully molten magma ocean, indicated by a melt fraction $\Phi_{\rm global} = 1$ (first column in Fig.\,\ref{figure grid scatter point}) and $T_{\rm surf} \gtrsim 3400$\,K (second column). To reproduce the observed density while retaining an atmosphere, our models suggest that TOI-561\,b may host a thick atmosphere, with surface pressures from $\sim$$10^{3}$ to a few $10^{4}$\,bar (third column). Such atmospheres are metal-rich with high mean molecular weights of $\sim$38--60\,\,g\,mol$^{-1}$, dominated by oxidized, high-mass species such as CO$_2$ (VMR$\sim$70\%) or SO$_2$ (VMR=70--90\%).

\subsubsection{Modeled atmospheric composition of TOI-561\,b}

\begin{figure*}
    \centering
    \includegraphics[width=\textwidth]{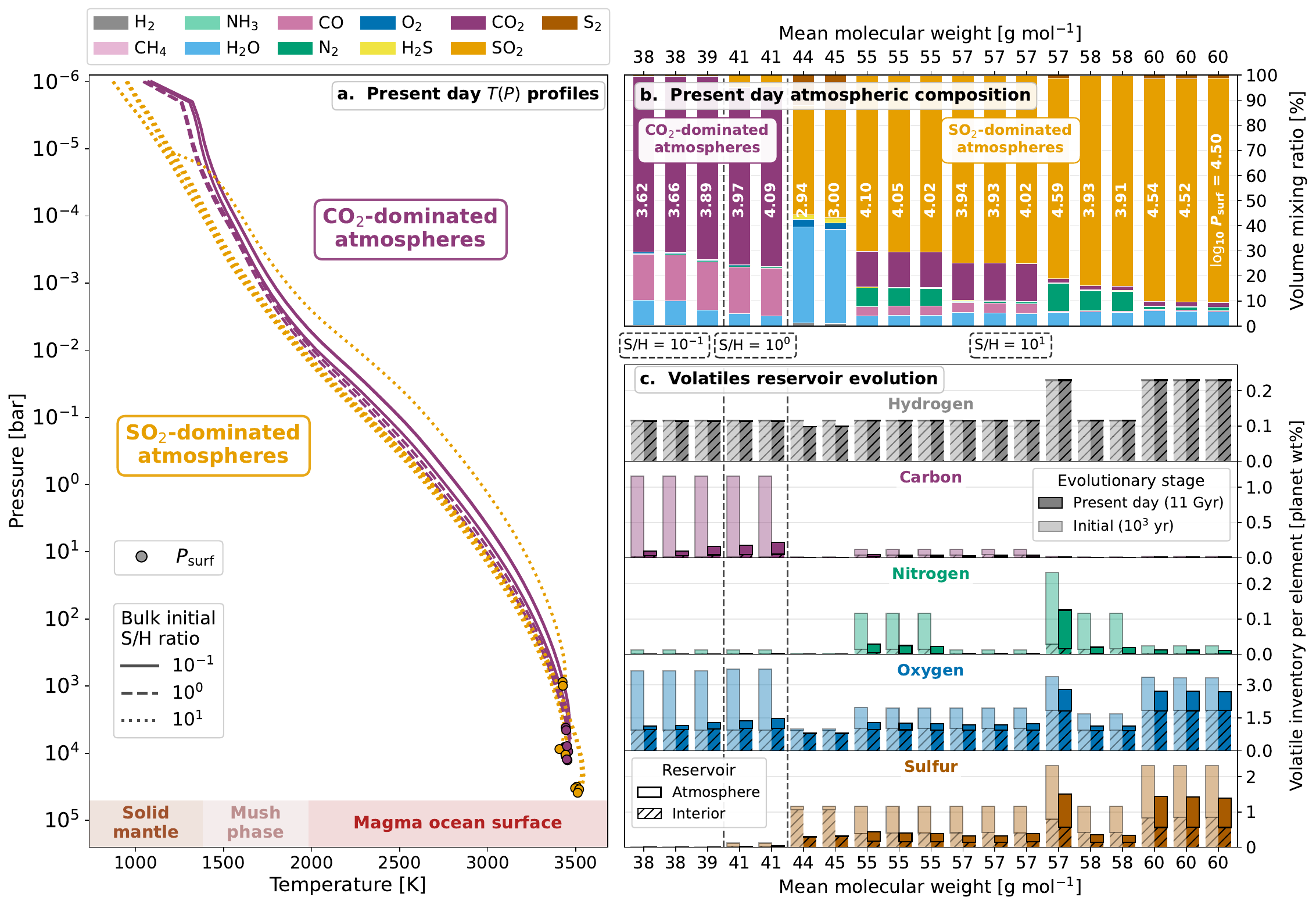}
    \caption{Modeled present-day properties of TOI-561\,b for the in-situ scenario~A, restricted to the simulations matching every observational constraint (bulk density, radius, mass and system age each within $1\sigma$). Panel\,a: Present-day radiative-convective atmospheric $T$($P$) profiles, one curve per case, colored by the dominant atmospheric species and styled by the bulk initial S/H ratio. Circles indicate the surface pressure. Panel\,b: Present-day atmospheric composition as volume mixing ratios, one stacked bar per case, ordered by mean molecular weight (top x-axis) and grouped by S/H ratio (dashed black lines). The $\log_{10} P_\mathrm{surf}$ value is indicated on each bar in white. Panel\,c: Volatile-reservoir evolution per element (C--H--O--N--S), organized by mean molecular weight (bottom x-axis) and stacked by S/H ratio (dashed black lines). We compare initial ($10^3$\,Gyr, shaded bars) with present day ($\sim$11\,Gyr, opaque bars) inventories. Each bar is split into the atmosphere (solid fill) and the interior (hatched), expressed as planet weight percent.}
    \label{plot compo atm}
\end{figure*}

\begin{figure*}
    \centering
    \includegraphics[width=\textwidth]{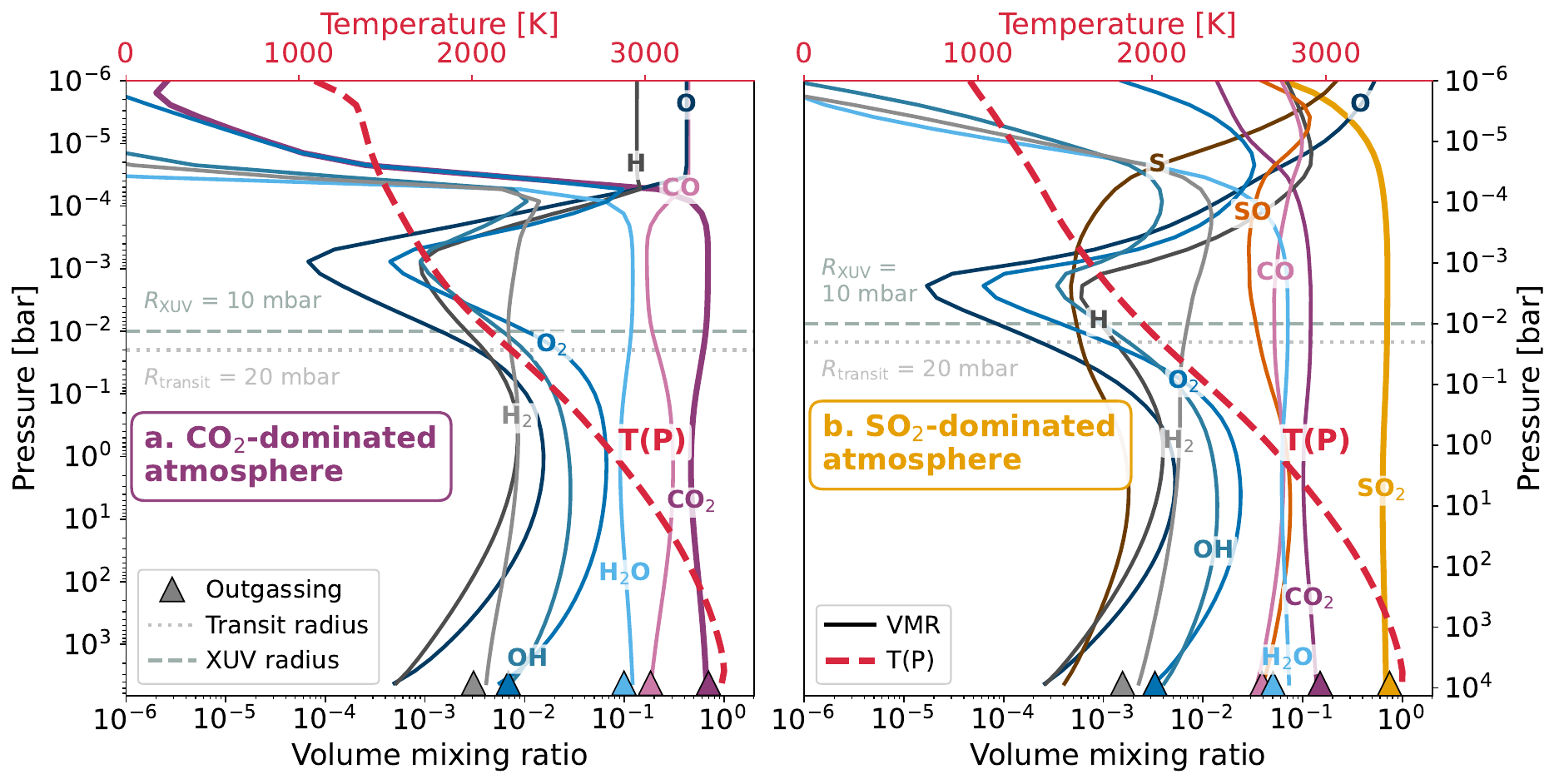}
    \caption{Vertical volume mixing ratios (solid lines) and $T$($P$) profiles (red dashed lines) of the two atmospheric archetypes (in-situ scenario A), CO$_2$-dominated (panel\,a) and SO$_2$-dominated (panel\,b), after post-processing photochemical computation with \texttt{VULCAN} \citep{Tsai_2021, Tsai_2023}. Solid lines show the photochemical mixing ratio of each major species as a function of pressure, colored by species. Triangles mark the outgassed abundances of the displayed species at the end of the \texttt{PROTEUS} run, i.e. before photochemistry.}
    \label{plot VMR2 archetypes}
\end{figure*}

Figure\,\ref{plot compo atm} summarizes the present-day atmospheres of the 19 simulations that reproduce all the observed bulk properties of TOI-561\,b for the in-situ scenario A. Present-day $T$($P$) profiles are shown in panel\,a, present day atmospheric composition in panel\,b, and the partitioning of each volatile element between the atmosphere and the interior from birth to present day in panel\,c. 

All retained atmospheres are thick, with surface pressures ranging from $10^{2.94}$ to $10^{4.59}$\,bar (Fig.\,\ref{plot compo atm}, panel\,b) overlying a global magma-ocean surface ($T_\mathrm{surf}\approx3410$--$3520$\,K, panel\,a), and high mean molecular weights of $38$--$60$\,g\,mol$^{-1}$ (panel\,b and c).

The present day $T$($P$) profiles (Fig.\,\ref{plot compo atm}, panel\,a) are uniform across all 19 valid cases for the in-situ scenario A. Each atmosphere allows for radiative or convective layers, spanning from a global magma-ocean surface ($T_\mathrm{surf}\approx3410$--$3520$\,K) up to ${\sim}1000$--$1500$\,K at the top of the modeled atmosphere. The S/H ratio (line style) and dominant species (color) do not have a strong impact on the profiles. Our profiles are consistent with the radiative-convective behavior found for molten rocky planets with volatile-rich atmospheres modeled by \citet{Nicholls_2025_MNRAS, nicholls_beyond_2026}. 

Panel\,b shows that two atmospheric archetypes emerge from our simulations, set by the bulk initial sulfur content: CO$_2$-dominated atmospheres for S/H\,$\leq1$ (5\,cases) and SO$_2$-dominated atmospheres for S/H\,$=10$ (14\,cases). 
For the carbon-rich archetype, CO$_2$ makes up ${\sim}70\%$ of the atmosphere, with CO as the second most abundant species (${\sim}19\%$) and a few percent of water ($4$--$10\%$). Carbon-rich outcomes are analogous to a super-Venus, with a broadly similar atmospheric composition but substantially higher surface pressures: $\sim$90\,bar on Venus \citep{Wood_1968}, compared to several thousand bar in our simulations. This scenario is thus more consistent with a puffy Venus interpretation of under-dense lava worlds by \citep{Peng_Valencia_2024}. On Venus, the carbon-rich atmosphere is maintained by volcanic outgassing \citep{Bullock_2001}, whereas in our models outgassing is more efficient, sourced directly from a global magma ocean rather than from discrete volcanic activity. Recent JWST observations of 55\,Cnc\,e suggested the presence of a CO$_2$/CO-rich secondary atmosphere sustained by exchange with an underlying magma ocean \citep{Hu_2024}. Such a finding is qualitatively consistent with the carbon-rich outcomes identified here for TOI-561\,b.
For the sulfur-rich archetype, SO$_2$ dominates (typically $70$--$90\%$), with either CO$_2$ or N$_2$ as the secondary species ($\lesssim15\%$) and water generally below ${\sim}6\%$. Nevertheless, two cases stand out as more water-rich (H$_2$O\,${\approx}\,37$--$38\%$), resulting in the lowest surface pressures ($\sim10^{3}$\,bar). Both cases exhibit a high initial sulfur inventory (S/H\,=\,10) over an oxidized mantle ($f$O$_2$\,=\,IW+4), with an initial volatile inventory poor in carbon and nitrogen (C/H\,=\,0.1 and N/H\,=\,0.015). With CO$_2$/CO and N$_2$ strongly suppressed, H$_2$O becomes a major outgassed species alongside SO$_2$ \citep[according to their solubility laws under oxidized conditions;][]{Gaillard_2022,SOSSI_2023}, which also accounts for their comparatively low surface pressures.
For SO$_2$-dominated atmospheres, such planets may be viewed as a volcanic super-Io analog, sharing an atmosphere dominated by sulfur-bearing species sustained by volcanic outgassing.
However, the analogy is compositional only: our simulations exhibit substantially higher surface pressures ($\sim10^{3}$--$10^{4.6}$\,bar) compared to Io's tenuous atmosphere \citep[$\sim10^{-8}$--$10^{-6}$\,bar;][]{Lellouch_1990}. Our modeled TOI-561\,b atmospheres are sustained by continuous outgassing over $\sim$11\,Gyr of evolution, while Io's thin atmosphere is replenished by episodic volcanic events \citep{Lellouch_2003}.
A thick SO$_2$ atmosphere over a molten super-Earth is a plausible end-state, consistent with the volatile-rich, sulfur-bearing evolution found by \citet{Nicholls_2025_Nature} for the molten super-Earth L\,98-59\,d. In our models, this archetype is sustained by vigorous outgassing from the underlying magma ocean, which continually resupplies sulfur to the atmosphere over the lifetime of the planet (panel\,c).

Panel\,c highlights the coupled evolution of interior and atmospheric volatile reservoirs. It traces how the competing influences of magma-ocean outgassing, which replenishes and sustains the atmosphere, and atmospheric escape, which removes volatiles to space, shape the distribution of each element from planet formation to the present day. Carbon and nitrogen reside almost entirely in the atmosphere, reflecting their low solubility in silicate melt at oxidizing conditions \citep{Lichtenberg_2021_JGRP, Bower_2022, Suer2023FrEaS, Shorttle2024ApJL}. Oxygen and sulfur are partitioned across both reservoirs. However, hydrogen stays largely sequestered in the mantle (${\sim}99\%$) from the stars of our simulations ($10^{3}$\,yr) to present day. Hydrogen is bound with oxygen as water, which is highly soluble in the molten silicate mantle \citep{Suer2023FrEaS, SOSSI_2023}. Atmospheric escape is the dominant volatile sink over the evolution of the planet. By present day it has removed most of the nitrogen (${\sim}80\%$), carbon (${\sim}70\%$) and sulfur (${\sim}70\%$) inventories. Hydrogen is essentially left untouched in the mantle (${\lesssim}2\%$ lost). Sulfur shows a clear interplay between replenishment through outgassing and loss via atmospheric escape. Following the $f$O$_2$-dependent sulfur solubility law of \cite{Gaillard_2022}, sulfur becomes less soluble in the silicate melt under oxidized conditions ($f$O$_2 \gtrsim$\,IW+3) and is therefore preferentially outgassed, producing sulfur-rich atmospheres. Atmospheric sulfur is continuously resupplied from the magma ocean, even if escape drives its atmospheric abundance below the birth value as the interior reservoir is progressively depleted. Oxygen persists in every atmosphere, pairing with carbon or sulfur to form CO$_2$ or SO$_2$.

Figure\,\ref{plot VMR2 archetypes} shows the vertical volume mixing ratio (VMR, solid lines) and $T$($P$) profiles (red dashed lines) of the two atmosphere archetypes, CO$_2$-dominated (panel\,a) and SO$_2$-dominated (panel\,b), after post-processing photochemistry using \texttt{VULCAN} \citep{Tsai_2021, Tsai_2023}. 
Both cases share the same core-radius fraction, Bond albedo, escape efficiency factor and oxygen fugacity but differ only in their initial bulk volatile inventories. Triangles mark the outgassed abundances of species at the end of the \texttt{PROTEUS} run, before photochemical computation.
Near the surface, the composition matches the bulk outgassed mixture (triangles), indicating that the deep atmosphere is largely in thermochemical equilibrium. High temperatures and pressures near the magma ocean surface keep reaction timescales short compared to vertical transport and photolysis \citep{Moses_2011,tsai_2017}, validating our assumption of a well-mixed composition set to the outgassed mixture during the evolution in this region. Photochemistry alters the profile only at pressures below $\sim10^{-4}$\,bar, so the dominant CO$_2$/SO$_2$ species (and the archetype distinction) are preserved throughout the bulk atmosphere.

The dominant species, i.e. CO$_2$ ($\sim$70\%) and SO$_2$ ($\sim$71\%), are well mixed through the deep, optically thick atmosphere. Toward lower pressures (i.e. in the upper atmosphere), photochemistry leads to dominant species being photo-dissociated. For the carbon-rich case, CO$_2$ is partly dissociated into CO and atomic O, while H$_2$O feeds OH, H and O$_2$. At pressure less than $\sim10^{-4}$\,bar, the CO$_2$-atmosphere archetype presents CO, O and H in significant proportion ($\sim$40\% for CO and O, and $\sim$13\% for H) relative to other species. 
For the sulfur-rich case, SO$_2$ breaks down into SO, atomic S and O at very low pressure ($\sim10^{-5}$\,bar). In both archetypes atomic oxygen becomes a major upper-atmosphere constituent (VMR\,$\gtrsim0.4$--$0.5$), reflecting the strongly oxidized, H-depleted nature of heavy, thick atmospheres for USP planets.

\subsection{Regimes of rapid atmospheric loss}

\begin{figure*}
    \centering
    \includegraphics[width=\textwidth]{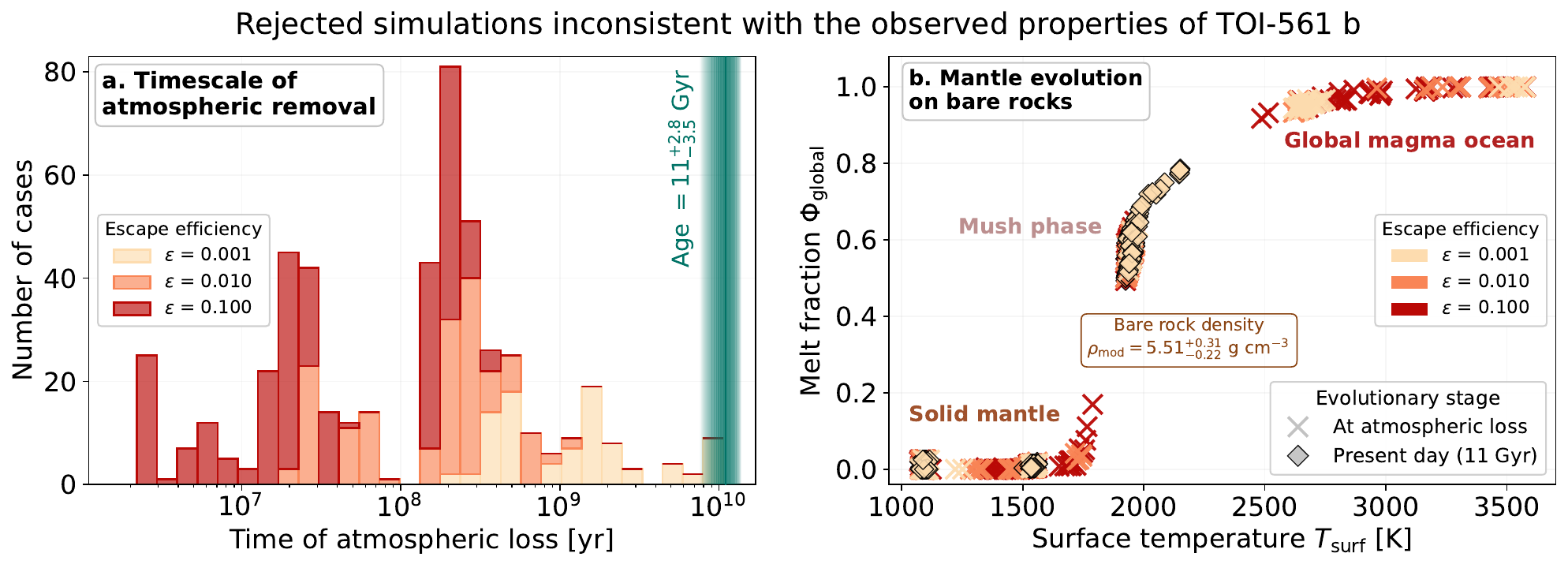}
    \caption{Timescale of atmospheric loss (panel\,a) and interior evolution (panel\,b) for the in-situ scenario A for simulations that fail to reproduce the bulk properties of TOI-561\,b, colored by escape efficiency $\epsilon$. Current system age is indicated by the green shaded region at $11^{+2.8}_{-3.5}$\,Gyr \citep{Lacedelli_2022} in panel\,a. In panel\,b, the interior state is shown at atmospheric loss (crosses) and present day (triangles). At both timescales, bare rocks reach a density of $\rho_{\rm mod}=5.51^{+0.31}_{-0.22}$\,g\,cm$^{-3}$ (mean value with extrema across all atmosphere-less simulations in the grid), far above the observed value.}
    \label{plot escape atm histogram}
\end{figure*}

Figure\,\ref{plot escape atm histogram} characterizes the in-situ (scenario\,A) simulations that fail to reproduce the observed bulk properties of TOI-561\,b due to atmospheric loss during their evolution. Panel\,a shows the time at which their atmosphere is lost. Panel\,b indicates the interior state at atmospheric removal (crosses) and at the present day system age (triangles), for the three tested escape efficiencies $\epsilon$.

Of the converged in-situ simulations, $\sim52\%$ lose their atmospheres (displayed in panel\,a in Fig.\,\ref{plot escape atm histogram}). Both the retention fraction and the removal timescale are governed primarily by the escape efficiency as shown in panel\,a. At low efficiency ($\epsilon=0.001$, beige histograms) the atmosphere survives the longest, being stripped between $10^8$ and $10^{10}$\,yr in only $20\%$ of cases. Higher escape efficiencies drive earlier atmospheric loss. For $\epsilon=0.01$ (orange) the atmosphere is removed between $10^7$ and $10^9$\,yr in $64\%$ of cases. For efficient escape regime with $\epsilon=0.1$ (dark red), the atmosphere is removed efficiently for $99\%$ of simulations at early evolutionary stages ($10^6$--$10^8$\,yr). The short orbital period of TOI-561\,b ($P=0.44$\,d) places the planet under intense XUV irradiation, so that for $\epsilon\ge0.01$ most atmospheres are eroded within the first Gyr. Above a mass loss rate of $\dot{M}_{\rm EL} = 10^{10}$\,g\,s$^{-1}$, none of our simulated planets retain an atmosphere.

At the time of atmospheric loss (crosses, panel\,b in Fig.\,\ref{plot escape atm histogram}), $24\%$ of our simulations are still in a global magma-ocean stage ($\Phi_{\rm global}>0.85$, $T_{\rm surf}\gtrsim2500$\,K). For those planets, a thick atmosphere insulates the surface and sustains a molten mantle. The other cases have already begun to crystallize by the time they lose their atmosphere. Once the atmosphere is removed, the bare rock cools down rapidly toward present-day. At 11\,Gyr, such planets fall into two regimes: molten at the surface but crystallized at depth ($39\%$), or with a fully solidified mantle ($61\%$).
A bare super-Earth at the orbit of TOI-561\,b can therefore remain partially molten after $11$\,Gyr, but cannot sustain a global magma ocean without an insulating atmosphere, according to our simulations. This is in line with previous multi-dimensional mantle convection simulations of USPs \citep{Meier2023AA,Meier2026MNRAS}, which focused on atmosphere-less or thin-atmosphere scenarios.
Our 1D model captures only the global-mean melt fraction, but on a tidally-locked bare USP planet like TOI-561\,b, this residual melt would likely form a dayside magma pool \citep{Leger_2011,Kite_2016}.
Regardless of their interior state, all atmosphere-less cases are bare rocky mantles with a present day bulk density $\rho_{\rm mod}=5.51^{+0.31}_{-0.22}$\,g\,cm$^{-3}$ (mean value with extrema across all atmosphere-less simulations), well above the observed $\rho_{\rm obs}=4.3049^{+0.4411}_{-0.4216}$\,g\,cm$^{-3}$ \citep{Patel_2023} for every case. 
Early atmosphere removal motivates the investigation of a late inward scenario B where the planet receives less XUV radiations, lowering escape rate and sequestrating volatiles on longer timescale in the interior, before reaching its present orbit. 

\subsection{Impact of orbital migration on atmospheric evolution}

\begin{figure*}
    \centering
    \includegraphics[width=\textwidth]{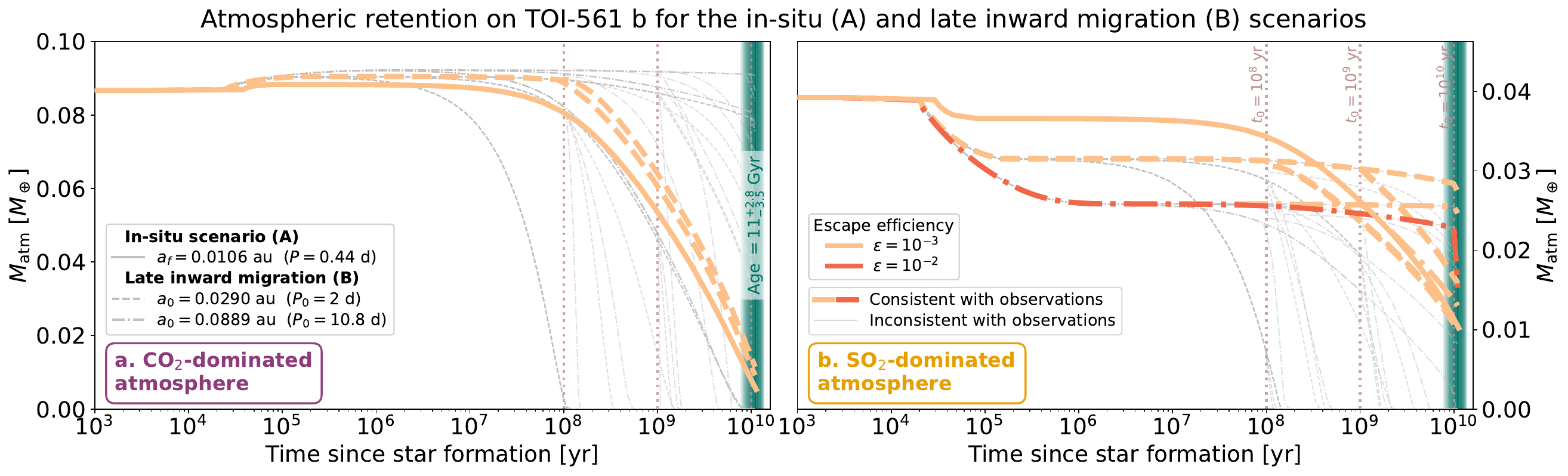}
    \caption{Atmospheric mass $M_{\rm atm}$ evolution of TOI-561\,b for CO$_2$-dominated (panel\,a) and SO$_2$-dominated (panel\,b) atmospheres. Solid lines represent the in-situ scenario A at a fixed semimajor axis of $a_0=a_{\rm f}=0.0106$\,au. Other curves show the late inward migration scenario B, with line style corresponding to initial orbital separation ($a_0=0.029$\,au dashed and $a_0=0.0889$\,au dash-dotted). Onset migration times $t_0$ are indicated with brown dotted verticals lines. The green region indicates the stellar age, $11^{+2.8}_{-3.5}$\,Gyr \citep{Lacedelli_2022}. Colored tracks (by escape efficiency $\epsilon$) match the observed density, radius and mass ($1\sigma$) within the age window with a circular present-day orbit ($e_{\rm f}\simeq0$), while gray tracks are inconsistent with observations.}
    \label{plot orbital migration}
\end{figure*}

Figure\,\ref{plot orbital migration} presents the atmospheric mass evolution when varying the escape efficiency factor $\epsilon$ for 2 archetypes atmospheres: CO$_2$- (panel\,a) and SO$_2$-dominated (panel\,b) atmosphere. We compare the in-situ scenario A (solid lines) with the late inward migration scenario B for 2 initial separations $a_0=0.029$\,au (dashed lines) and $a_0=0.0889$\,au (dash-dotted lines). These correspond respectively to initial periods $P_0=2$\,days and $P_0=10.8$\,days. We choose them to span the low-order resonances with the inner companion c, consistent with a resonant-chain origin (Sect.\,\ref{section scenario a and b in methods}). We display the evolution of semimajor axis and eccentricity for a single case in the Appendix, for more details see Fig.\,\ref{plot appendix sma ecc evolution}.

Of the 76 converged migration simulations, 34 retain an atmosphere at present-day ($P_{\rm surf}>0$), but only $\sim15\%$ (11 cases) also reproduce the observed radius, density and mass within $1\sigma$ at the system age on a circularized orbit.
Across all orbital histories the surviving atmospheres remain either CO$_2$-dominated (79–81\% by VMR) or SO$_2$-dominated (59–75\%) across the explored parameter space. This reflects the successful atmospheric retention cases selected from scenario A as initial conditions for the migration models. Only two cases with a CO$_2$-dominated atmosphere are valid when considering orbital migration against nine for the SO$_2$-archetype. Although carbon-rich atmospheres start roughly twice as massive ($M_{\rm atm}\sim0.09\,M_\oplus$ vs.\ $\sim0.04\,M_\oplus$), their larger scale height drives more efficient escape, leaving comparable present-day masses ($\sim0.01\,M_\oplus$ for CO$_2$; $0.01$–$0.03\,M_\oplus$ for SO$_2$).

At orbital distances larger than the present day value ($a_{\rm f}=0.0106$\,au), weaker XUV irradiation reduces atmospheric escape rates and favors retention relative to the in-situ case (solid lines). Only one case survives with a moderate escape efficiency ($\epsilon=10^{-2}$) with the latest onset migration time during the last Gyr of evolution ($t_0=10^{10}$\,yr, orange dash-dotted line in panel\,b). All other consistent cases share the low escape efficiency $\epsilon=10^{-3}$ of the in-situ scenario A.

The migration parameters are only weakly constrained. The two CO$_2$-consistent cases share an early onset ($t_0=10^8$\,yr) with fast-to-intermediate migration timescales ($\tau=10^6$–$10^8$\,yr; not encoded in Fig.\,\ref{plot orbital migration} for clarity). For the SO$_2$ archetype, every onset time reproduces the observations, but the slowest migration ($\tau=10^{10}$\,yr) never does, favoring $\tau\lesssim10^8$\,yr in this case.
Cases starting at a shorter period $P_0=2$\,days ($a_0 = 0.029$\,au, dashed lines) are more plausible than the ones starting at 10.8 days ($a_0 = 0.0889$\,au, dashed-dotted lines), accounting for 8 of the 11 observationally-consistent tracks. For an initial orbit at the present day period of planet c ($P_0=10.8$\,days), only three SO$_2$-dominated cases survive and none of the CO$_2$-dominated atmospheres. This wider-orbit corner is also the least physically favorable: reaching the present-day orbit from $a_0=0.0889$\,au requires exciting a large initial eccentricity ($e_0\approx0.94$), whose feasibility and small-periapsis implications we discuss in Sect.\,\ref{section model limitations}. Both the low number of successful high-$a_0$ tracks and their weak dynamical justification disfavor the extreme-eccentricity pathway, supporting an inward migration onset near $P_0=2$\,days.

Both the in-situ A and late inward migration B pathways thus remain viable to explain present day TOI-561\,b properties.

\section{Discussion}
\label{section discussion}

\subsection{Model assumptions and limitations}
\label{section model limitations}

We discuss here the assumptions and limitations of our coupled interior--atmosphere modeling with \texttt{PROTEUS} and their impact on our main findings. The main limitations are the uncertainty in the stellar and escape history, the neglect of tidal heating and orbital survival in scenario\,B, and our 1D global-mean treatment. The remaining assumptions (post-processed photochemistry, clear-sky radiative transfer, the neglect of rock vapor, and equilibrium outgassing) affect specific quantities but not our main conclusions.

Atmospheric escape is integrated over $\sim11$\,Gyr using \texttt{MORS} stellar evolution tracks \citep{Johnstone_2021}. The system age is clearly older than the Sun, but still somewhat uncertain \citep[7.5--13.8\,Gyr;][]{Lacedelli_2022} and because TOI-561 is an old, metal-poor thick-disk star, its rotation and activity history are poorly constrained. Since atmospheric retention in our models is set largely by the XUV irradiation, stellar evolution represents a major caveat. We assume a constant escape efficiency $\epsilon$, although it depends on planetary mass, radius, and incident flux \citep{Owen-2012} and generally
decreases as an atmosphere thins. Our mass-loss rates are therefore conservative upper limits, and $\epsilon \lesssim 10^{-3}$ should be read as an effective value averaged over the evolution.

All simulations are computed under the assumption of monotonic atmospheric cooling. This assumption is well justified for the in-situ configuration (scenario~A), in which the instellation is constant and the planet cools secularly. In scenario\,B, the increasing instellation during inward migration can drive net warming, and our treatment therefore yields conservative lower limits on the scenario\,B surface temperatures. Because these cases already correspond to global magma oceans, this limitation does not affect our conclusions.

Our late inward migration scenario B involves a high eccentricity phase, which raises additional caveats regarding atmosphere survival. First, exciting the eccentricity to $e\approx0.80$--$0.94$ is dynamically demanding. The known planet companions \citep[TOI-561\,c, d, e;][]{Lacedelli_2022} are close-in and low-mass, so alternative routes such as planet--planet scattering \citep{Carrera2019scattering} or a stellar flyby may be required \citep[e.g.,][]{Rasio_1996,Chatterjee_2008}. Second, at peak eccentricity, the periapsis is $\approx0.0055$\,au ($\approx1.4\,R_\star$), between the rigid and fluid Roche limits, so survival against tidal disruption is not guaranteed, especially for the longest migration timescales ($\tau\sim10^{10}$\,yr, Fig.\,\ref{plot orbital migration}). Third, our orbit-averaged XUV flux underestimates escape, since the high-flux passages occur near periapsis. 

We do not model tidal heating. All our successful cases are already global magma oceans at present day ($\Phi_{\rm global}=1$ in Fig.\,\ref{figure grid scatter point}), sustained by strong instellation and the insulating thick atmosphere. For in-situ scenario A, tidal heating would only prolong this molten state \citep{Farhat2025ApJ,Nicholls_2025_tidal_l9869}, not affecting our main conclusions. In scenario\,B however tidal heating during the high-eccentricity phase is far larger. For a molten mantle, the orbit-averaged rate reaches $\sim1.1\times10^{22}$\,W, comparable to the $\sim5\times10^{21}$\,W of instellation absorbed at periapsis by the planet in our simulations. Near periapsis, it could inflate the atmosphere \citep{Millholland2019} and enhance escape \citep{Barnes2013} rather than only prolong the magma-ocean epoch. We therefore present scenario B as a plausibility demonstration rather than a robust prediction of atmospheric survival.

We treat the planet as a single global-mean column. However, TOI-561\,b is a tidally locked USP ($P=0.44$\,d) planet, so its dayside and nightside can differ substantially -- a contrast that a thick atmosphere would reduce but that a 1D model cannot capture \citep{Castan_2011,Kite_2016}. Thus, our modeled quantities such as $P_\mathrm{surf}$, $T_\mathrm{surf}$, and escape rates represent global average quantities. In our late inward migration scenarios (B), volatiles could be cold-trapped on the cold nightside \citep{Castan_2011,Kite_2016,Nguyen_2020}, setting up a day-night volatile cycle. However, as all our successful simulations retain thick envelopes, this would only impact the upper atmosphere and not substantially feed back onto the surface or interior conditions. A deep, long-lived magma ocean beneath the nightside surface could recycle these trapped volatiles and resupply them from the interior \citep{Dorn_lichtenberg_2021, Meier2024JGRE, Meier2026MNRAS, Lichtenberg_2025}. Lateral transport could spread asymmetrically outgassed volatiles across the planet \citep{Nguyen_2020} and buffer the atmosphere against nightside condensation and dayside escape. Thus, nightside cold-trapping and interior resupply buffer the volatile envelope of TOI-561\,b, while the enhanced day--night contrast drives dayside escape. Their net effect on atmospheric stability cannot be determined from our 1D treatment.

We apply photochemistry \citep[with \texttt{VULCAN;}][]{Tsai_2021,Tsai_2023} only at present day, not throughout the evolution. Because the photochemical profiles (solid lines in Fig.\,\ref{plot VMR2 archetypes}) are post-processed on the converged radiative–convective $T$($P$) profiles from \texttt{AGNI} (red dashed lines in Fig.\,\ref{plot VMR2 archetypes}), the photochemically altered composition does not feed back onto the thermal structure, although this feedback is generally modest for irradiated, volatile-rich atmospheres \citep{Drummond_2016, Agundez_2025,Panagiotou_2026}. 
The post-processed photochemistry is driven by a PHOENIX synthetic spectrum \citep{Husser_2013_phoenix}, which lacks the chromospheric and coronal emission (X-ray, EUV, Ly$\alpha$) that dominates a star's high-energy output \citep{France_2016}. Thus our resulting computations for photodissociation represent a lower limit. Atmospheric escape is unaffected, as its XUV flux is taken from the MORS rotation tracks \citep{Johnstone_2021}, not the synthetic spectrum.
The present-day VMR profiles show a strongly dissociated upper atmosphere. Above $\sim10^{-3}$\,bar (for a CO$_2$ atmosphere) and $\sim10^{-5}$\,bar (for a SO$_2$ case), CO$_2$, H$_2$O and SO$_2$ break down into CO, O, OH, H, SO and S. Treated self-consistently during the evolution, this photochemistry could also form hazes and aerosols that alter the upper-atmosphere thermal structure  \citep[Fig.\,\ref{plot compo atm}, panel a;][]{Hu_2012, Tsai_2021, kitzmann_2023}.

Our model assumes a clear-sky atmosphere. Condensation and rainout are permitted in the model but do not occur in any of the valid scenario A cases, as the two dominant species (CO$_2$ and SO$_2$) remain supercritical at the surface pressures reached here ($\sim10^3$--$10^4$\,bar) and the atmospheres stay above their saturation temperatures at all levels ($T \gtrsim 880$\,K throughout). Clouds and hazes are not explicitly modeled here although they can originate from the photochemical hazes noted above or from condensates of species outside our inventory (e.g. rock vapor). Such a layer could modify the radiative budget by increasing the albedo, as inferred from JWST observations \citep{Boucher_2026}, or by adding a greenhouse warming \citep{2023selsis} — but resolving such effects is beyond the scope of our clear-sky treatment.

Our atmospheres contain only volatiles (C--H--O--N--S) and neglect rock vapor species (SiO, Na, K, Fe, Mg), which are expected at our present-day surface temperatures \citep[$T_\mathrm{surf}\approx3410$--$3520$\,K;][]{Piette_2023,Janssen_2026}. Including vaporized rock species would raise the atmosphere mean molecular weight, add opacity, and create a strong thermal inversion in the upper atmosphere for modeled $T$($P$) structures of TOI-561\,b \citep{Zilinskas_2022,Janssen_2026}. Our CO$_2$- and SO$_2$-dominated archetypes thus represent upper limits on volatile content, and may be partly diluted by vaporized rocks.

Outgassing assumes thermochemical equilibrium at the surface (\texttt{CALLIOPE}), at a mantle oxygen fugacity ($f$O$_2$) that follows the temperature-dependent IW buffer \citep{ONEILL_2002} at a fixed offset ($\Delta$IW), rather than evolving with mantle crystallization or core formation. The solubility treatment also models sulfur as sulfide (S$^{2-}$) only. Under the oxidizing conditions favored here, sulfur also dissolves as sulfate (S$^{6+}$), whose saturation content rises steeply with $f$O$_2$ \citep{Jugo_2009, Boulliung_2022}. We therefore under-retain sulfur in the magma ocean, so the outgassed sulfur (and thus the SO$_2$-dominated archetype) should be read as an upper limit.

The observationally consistent region is bounded by two grid edges simultaneously as shown in Fig.\,\ref{figure grid scatter point}: an oxidized mantle ($f$O$_2 \gtrsim \mathrm{IW}+4$), a low escape efficiency ($\epsilon \lesssim 0.001$), and an initial hydrogen inventory (H\,$\lesssim 200$\,EO). Our analysis therefore constrains these parameters only one-sidedly, and the true joint optimum may lie outside the sampled parameter box. Likewise, fixing the initial mass to the present-day value (Sect.\,\ref{section methods proteus}) excludes an evaporated-core origin after substantial primordial H/He loss. Although disfavored by the low observed density \citep[$\rho_{\mathrm{obs}}\,=\,4.3049^{+0.4411}_{-0.4216}$\,g\,cm$^{-3}$;][]{Patel_2023}, this leaves a region of parameter space unexplored.

\subsection{Physical interpretation of the surviving atmospheres}
\label{section discussion physical atmosphere}

We model atmospheric escape using an energy-limited formalism \citep[][see Sect.\,\ref{section methods proteus}]{Watson_1981, Erkaev_2007, lopez-fortney-2013}. Following \citet{Nicholls_2025_Nature}, we assume that all escaping species are lost in atomic form (C--H--N--S), consistent with a photodissociated upper atmosphere shown in Fig.\,\ref{plot VMR2 archetypes}. Oxygen is the exception; rather than tracking it as an escaping element with a cumulative loss, we recompute its atmospheric abundance at each timestep, as set by the mantle redox state ($f$O$_2$) and buffered by the large interior oxygen reservoir (panel\,c in Fig.\,\ref{plot compo atm}). The total escape rate is partitioned among elements according to the atmospheric elemental mass mixing ratios derived from the outgassing calculations (triangles in Fig.\,\ref{plot VMR2 archetypes}). These elemental ratios are conserved through the column (mixing and photodissociation redistribute species but not elemental abundances) so the outgassed values are representative of the material available to escape.

We adopt low escape efficiencies, $\epsilon = 0.001$--$0.1$, below the $\epsilon=0.15$ often used for modeling super-Earths and rocky planets \citep{Kasting-1983, Luger_Barnes_2015, Erkaev-2016, Kubyshkina-2018, Owen_2019, Cherubim_2024}. Such low escape efficiencies remain appropriate for our metal-rich, high MMW atmospheres (CO$_2$/SO$_2$, MMW\,$\approx38$--$60$\,g\,mol$^{-1}$), for two dependent reasons. First, a high MMW atmosphere has a small scale height and hence a small $R_\mathrm{XUV}$. Since $\dot{M}_{\rm EL}\propto R_\mathrm{XUV}^3$, the mass loss rate is strongly reduced regardless of $\epsilon$ \citep[as proposed by][for water-rich atmospheres]{Yoshida-gaidos-2025}. Second, the heating efficiency factor $\epsilon$ itself is lower in metal-rich gas, because atomic line emission and radiative cooling (especially for C bearing species) reduce the total XUV energy available to drive a hydrodynamic outflow \citep{Shematovich-2014, Nakayama-2022, Yoshida-2024, Chatterjee_2026}. These studies set the direction of this effect but not its magnitude, making our adopted $\epsilon=10^{-1}$--$10^{-3}$ physically motivated rather than quantitatively predicted.

Moreover, our modeled atmospheres are strongly hydrogen-depleted (H$_2$O\,$<10\%$; Fig.\,\ref{plot compo atm}, panel\,b), a prediction that a relatively flat transmission spectrum in upcoming work could confirm. With no light H background to preferentially drag the heavier species (C, N, S), fractionating escape is negligible, consistent with our model treatment. The intense XUV irradiation on this USP orbit ($P=0.44$\,days) drives a strong, collisional outflow that drags all species off together, keeping the escape essentially unfractionated.

The observed low density of TOI-561\,b cannot be reproduced by a bare solid interior, which is too dense (Fig.\,\ref{plot escape atm histogram}, panel\,b). As suggested by previous studies \citep{teske_2025,Boucher_2026}, a retained volatile envelope is required to match observed density as suggested by our model (Fig.\,\ref{figure grid scatter point}). Our successful cases thus describe a present-day state with a thick, heavy CO$_2$- or SO$_2$-dominated atmosphere overlying a global magma ocean. 

An alternative our models cannot capture is a long-lived, helium-enriched envelope, left behind when hydrogen escapes preferentially over the heavier helium -- a scenario predicted for planets near the radius valley, including TOI-561\,b \citep{Hu_2015, Malsky_2023, Cherubim_2024, Cherubim_2025}. Such fractionation may be more likely at the lower XUV fluxes of our scenario B with wider initial orbit. Because \texttt{PROTEUS} does not include helium, assessing this scenario though its low mean molecular weight ($\sim$4\,g\,mol$^{-1}$) would distinguish it from our high MMW atmospheres (38--60\,g\,mol$^{-1}$). However, an helium-rich envelope seems disfavored by the observed density \citep[$\rho_{\mathrm{obs}} = 4.3049^{+0.4411}_{-0.4216}$\,g\,cm$^{-3}$;][]{Patel_2023}.

\subsection{Interior geophysical state of TOI-561\,b}
\label{section discussion interior}

TOI-561\,b orbits very close to its host star ($P=0.44$\,d) and receives $\sim3900$ times the present-day instellation of Earth, that keeps its surface molten. In the cases that match the observational constraints, the planet hosts a global magma ocean, with melt fraction $\Phi_\mathrm{global}=1$ and surface temperature $T_\mathrm{surf}\gtrsim3400$\,K (Fig.\,\ref{figure grid scatter point}, columns 1--2). TOI-561\,b is therefore best described as a lava world by our model: a fully molten mantle beneath a thick, high MMW atmosphere.

These cases also favor a small iron core, with a core-radius fraction CRF\,$=0.3$--$0.4$ (Fig.\,\ref{figure grid scatter point}, panel\,5), smaller than Earth's. A smaller iron core implies a larger silicate mantle, which acts as a deep reservoir that stores volatiles dissolved in the melt over geological timescales \citep{Dorn_lichtenberg_2021,Moore_2023,Lichtenberg_2025}. In our models, hydrogen in particular remains almost entirely mantle-bound (paired with oxygen as dissolved OH$^{-}$ ions, panel\,c in Fig.\,\ref{plot compo atm}), buffering the surface volatile budget. This reservoir continuously outgasses and replenishes the atmosphere over the planet's evolution. Sustained outgassing acts as the principal counterbalance to atmospheric escape, favoring TOI-561\,b to retain an atmosphere over the system lifetime.

\subsection{Comparison with JWST observations and spectral predictions}

Two recent JWST/NIRSpec studies \citep{teske_2025,Boucher_2026} both find that TOI-561\,b retains a thick atmosphere to match observed density, consistent with our findings (Fig.\,\ref{figure grid scatter point}). \citet{teske_2025} showed that TOI-561\,b dayside emission spectrum is inconsistent with a bare rock at high significance, requiring a thick volatile envelope. Using a full-orbit phase curve, \cite{Boucher_2026} reach the same conclusion and infer a high Bond albedo as heat redistribution alone cannot explain the observed phase curve for TOI-561\,b. Both works attribute the atmosphere to volatile exchange with the interior -- precisely the magma-ocean reservoir that sustains the atmosphere in our models (Sect.\,\ref{section discussion interior}). Our simulations independently support the suggestion of a retained envelope.

Our simulations produce two atmospheric archetypes: either a CO$_2$-dominated (super-Venus like) or SO$_2$-dominated (super-Io like) that could explain a volatile envelope for TOI-561\,b.
The photospheric radius probes the $\sim$20\,mbar level (Fig.\,\ref{plot VMR2 archetypes}), well above the surface but below the upper layers where photochemistry strongly dissociates the gas into atomic species. At transit radius level the volume mixing ratios of the dominant outgassed species are essentially unchanged from the surface: CO$_2$ ($\sim$63\%) and SO$_2$ ($\sim$70\%) remain dominant.
The CO$_2$-dominated, super-Venus cases have their principal feature near $4.3\,\mu$m, while the SO$_2$-dominated, super-Io cases feature near $4\,\mu$m. A tentative excess at $4.3\,\mu$m presented by \citet{teske_2025} for the emission spectrum, and more prominently in the phase curve analysis of \cite{Boucher_2026}, could be consistent with our CO$_2$ archetype (though producing an emission rather than absorption feature likely requires clouds and/or rock vapor to drive a thermal inversion), but it is not statistically significant. The current data show no clear SO$_2$ signature. A detailed comparison of synthetic archetype spectra with the JWST data, including cloud effects, is deferred to upcoming work.

Reproducing TOI-561\,b's observed spectrum in detail likely requires clouds, which we do not model (Sect.\,\ref{section model limitations}). Our cloud-free CO$_2$ atmospheres have a largely non-inverted thermal structure except for a radiative near-surface layer (Fig.\,\ref{plot compo atm}, panel\,a), predicting a strong CO$_2$ absorption feature rather than a small emission feature at $4.3\,\mu$m. A high cloud deck would raise the photosphere level and flatten the temperature contrast across the band. This would weaken the feature toward a lower amplitude closer to the observed emission spectrum \citep{teske_2025}. Clouds would also raise the Bond albedo, reconciling our clear-sky models with the highly reflective atmosphere inferred by \cite{Boucher_2026}.

\subsection{Formation pathways of volatile-rich super-Earths}
\label{section discussion formation}

To match the observed low density, our models require TOI-561\,b to have formed volatile-rich (with an initial hydrogen content of H\,$\le200$\,EO and initial sulfur abundance as S/H\,$\le10$), retaining enough volatiles to sustain a thick atmosphere over its $\sim11$\,Gyr lifetime. A similarly volatile-rich formation was proposed by \citet{Nicholls_2025_Nature} for the molten super-Earth L\,98-59\,d (a more temperate, non-USP planet orbiting an M dwarf), in agreement with our results. This is notable because ultra-short-period planets are generally expected to be rocky and volatile-poor, having lost their primordial atmospheres to intense stellar irradiation \citep{Winn_2018}. A volatile-rich atmosphere TOI-561\,b therefore implies a large initial volatile budget and efficient retention against escape (Sect.\,\ref{section discussion physical atmosphere}).

Our successful cases favor an oxidized mantle with an initial oxygen fugacity at least $f$O$_2\ge$\,IW$+4$ (Fig.\,\ref{figure grid scatter point}, panel\,20). Two pathways could produce such an oxidized, volatile-rich composition. First, the planet could have accreted oxidized, ice-rich material -- for example beyond the water snowline -- and later migrated inward \citep{Lichtenberg2019NatAs,Bitsch2019AA,Lambrechts2019AA}. Such material delivers both volatiles and oxygen \citep{Doyle_2019, Ortenzi_2020}, while later disk migration brings the planet to its present USP orbit \citep{Carrera_2019,brandenberger_2026}. Our volatile-rich and oxidized solutions thus favor a migration history for the USP TOI-561\,b \citep{Krijt2023ASPC,Drazkowska2023ASPC}, and contrasts with the in-situ origin proposed for some USP planets \citep{Batygin2023NatAs}. 
Second, TOI-561\,b's mantle could have been oxidized internally by iron disproportionation during the planet's evolution \citep{Wade_2005,Armstrong_2019}. In the interiors of rocky planets, and especially of massive super-Earths, high mantle pressures drive the reaction $3\,\mathrm{FeO}\rightarrow\mathrm{Fe}+\mathrm{Fe_2O_3}$. The metallic iron (Fe) is denser and sinks to the planetary core, leaving ferric iron (Fe$_2$O$_3$) behind in the mantle and raising its oxygen fugacity \citep{Frost_2004, Wade_2005, Armstrong_2019, Hirschmann_2022, zhang2024ferric}. Earth illustrates this: reducing in bulk, yet with an upper mantle oxidized to $\approx\mathrm{IW}+4$ \citep{Frost_McCammon_2008}, comparable to the value favored by our successful cases. Because TOI-561\,b is more massive than Earth, this mechanism could in principle oxidize its mantle even more efficiently \citep{Frost_McCammon_2008,Hirschmann_2022}. However, it has been proposed that this oxidation pathway breaks down at super-Earth and sub-Neptune sizes. One route is a redox hysteresis, in which entrainment of metallic iron in the vigorously convecting magma ocean suppresses its rainout to the core \citep{Lichtenberg2021ApJL}. Another is a pressure-induced high-spin$\rightarrow$low-spin transition of ferrous iron stabilizes Fe$^{2+}$ at super-Earth interior pressures instead of driving its disproportionation, halting---and potentially reversing---the increase in mantle Fe$^{3+}$/$\Sigma$Fe (and hence $f$O$_2$) otherwise expected at higher pressures \citep{Girani2026EPSL}. If the oxidation of TOI-561 b is due to an endogenous mechanism instead of exogenously accreted materials, then the latter two mechanisms are disfavored by our results.

Oxidation alone does not require orbital migration, since disproportionation can oxidize the mantle in situ. The two pathways are not exclusive: disproportionation could oxidize the mantle internally while migration delivers volatiles from beyond the snowline. Either way, it is the volatile enrichment, not the oxidation, that points to substantial volatile accretion of TOI-561 b during formation and a possible history of inward-migration.

\section{Conclusions}
\label{section conclusion}

We investigated plausible evolutionary pathways for the USP super-Earth TOI-561\,b to match present-day observations, including density, mass and radius at inferred age of $11^{+2.8}_{-3.5}$\,Gyr \citep{Lacedelli_2022}. Using the coupled interior-atmosphere \texttt{PROTEUS} framework to model the evolution of TOI-561\,b over $\sim11$\,Gyr, we explored a wide parameter space toward atmospheric retention for an in-situ scenario A and a late migration scenario B. Our key findings are:

\begin{itemize}
    \item In agreement with \citet{teske_2025} and \cite{Boucher_2026}, our model predicts that TOI-561\,b needs to retain a massive atmosphere to match its present-day density \citep[$\rho_{\mathrm{obs}}\,=\,4.3049^{+0.4411}_{-0.4216}$\,g\,cm$^{-3}$;][]{Patel_2023}. A bare rock interior ($\rho_{\rm mod}=5.51^{+0.31}_{-0.22}$\,g\,cm$^{-3}$) is far too dense for every atmosphere-less case, within the range of core radius fraction we tested in this work (CRF $\geq0.3$).
    \item For the in-situ scenario A, our model suggests that TOI-561\,b started very oxidized ($f$O$_2$\,$\gtrsim$\,IW+4) and volatile-rich. Initial volatile compositions matching present day are $\le$\,200 Earth oceans of hydrogen, S/H\,$\le10$, and N/H\,$\le1$. To maintain an atmosphere over 11\,Gyr, the escape efficiency factor in the hydrodynamic escape regime must be low, with $\epsilon \lesssim 10^{-3}$, pointing to efficient line cooling.
    \item For all evolutionary pathways investigated, TOI-561\,b is consistent with a global magma ocean ($\Phi_{\rm global}=1$) below a thick ($P_{\rm surf} = 10^{3}$--$10^{4}$\,bar), heavy and high MMW atmosphere ($38$--$60$\,g\,mol$^{-1}$) at present day. 
    \item Two atmosphere archetypes emerge for TOI-561\,b after 11\,Gyr of evolution in our models, set by the initial bulk S/H ratio: CO$_2$-dominated and SO$_2$-dominated atmospheres.
    \item The late inward migration scenario B can also reproduce the present-day properties of TOI-561\,b, but only for short-to-intermediate migration timescales ($\tau = 10^{6}$--$10^{8}$\,yr). A migration onset near a $\sim$2-day period ($a_0 = 0.0290$\,au) is favored over a wider initial orbit ($a_0 = 0.0889$\,au).
    \item Both the in-situ (A) and late inward migration (B) pathways remain viable and require the same low escape efficiency ($\epsilon \lesssim 10^{-3}$), leaving the two histories observationally indistinguishable from present-day bulk properties alone.
\end{itemize}

Whether TOI-561\,b is ultimately a super-Venus (CO$_2$-rich) or a super-Io (SO$_2$-rich) remains a key open question. Answering it would make this old molten super-Earth a decisive benchmark for how volatile-rich rocky worlds evolve on ultra-short-period orbits.

\section*{Code and data availability}

\texttt{PROTEUS} is open-source software, available on GitHub at \url{https://github.com/FormingWorlds/PROTEUS}, together with its coupled submodules (\texttt{SPIDER}, \texttt{CALLIOPE}, \texttt{AGNI}, \texttt{SOCRATES}, \texttt{VULCAN}, \texttt{ZEPHYRUS}, and \texttt{MORS}). Extensive documentation for the coupled interior--atmosphere framework is provided at \url{https://proteus-framework.org}. 

The version of \texttt{PROTEUS} (v\,\texttt{25.11.19}; including orbital migration) used in this work, together with the simulated data and plots, is publicly available on Zenodo (DOI:10.5281/zenodo.22047651).

\software{\texttt{PROTEUS} \citep{Lichtenberg2021ApJL,Nicholls_2024_JGRP,Nicholls_2025_tidal_l9869, Nicholls_2025_MNRAS, Calder2026MNRAS,Sastre2026arXiv,vandijk_2026},
          \texttt{SPIDER} \citep{Bower_2018, Bower_2019, Bower_2022},
          \texttt{CALLIOPE} \citep{Bower_2022, SOSSI_2023,Nicholls_2024_JGRP,Nicholls_2025_MNRAS},
          \texttt{AGNI} \citep{Nicholls_agni_2025, Nicholls_2025_MNRAS},
          \texttt{SOCRATES} \citep{edwards_studies_1996, amundsen_radiation_2014, Sergeev_2023, Manners_2024},
          \texttt{VULCAN} \citep{Tsai_2021, Tsai_2023},
          \texttt{ZEPHYRUS} \citep{Postolec_2026},
          \texttt{MORS} \citep{Johnstone_2021},
          NumPy, Matplotlib, pandas, xarray}

\begin{acknowledgments}

Funded by the European Union (ERC, MagmaWorlds, 101219807). Views and opinions expressed are however those of the author(s) only and do not necessarily reflect those of the European Union or the European Research Council. Neither the European Union nor the granting authority can be held responsible for them. This research was further supported by the Branco Weiss Foundation, the Alfred P. Sloan Foundation (AEThER, G-2025-25284), NASA’s Nexus for Exoplanet System Science research coordination network (Alien Earths, 80NSSC21K0593), and the NWO NWA-ORC PRELIFE Consortium (NWA.1630.23.013). Support for US investigators (N.W. and J.T.) in program \#3860 was provided by NASA through a grant from the Space Telescope Science Institute, which is operated by the Association of Universities for Research in Astronomy, Inc., under NASA contract NAS5-03127. Support for program 3860 was provided by the Canadian Space Agency under contract 23JWGO2B06. This work has been partially funded by the Natural Sciences and Engineering Research Council of Canada (grant RGPIN-2021-02706).

H.N. acknowledges support from STFC grant UKRI1184.

L.D. and S.B. acknowledge support from the Natural Sciences and Engineering Research Council (NSERC), the Trottier Family Foundation and the Waterloo Centre for Astrophysics.

M.A. is supported by the Swiss National Science Foundation through the Postdoc.Mobility fellowship, grant number 230229. 

A.P. acknowledges funding from a UK Science and Technology Facilities Council (STFC) Small Award, grant number UKRI/ST/B001171/1.

We thank Mike Greklek-McKeon for insightful discussion about orbital migration. We thank the Center for Information Technology of the University of Groningen for providing access to the Hábrók high performance computing cluster. We made use of the Claude Code Command-Line Interpreter (Anthropic, 2026) for analysis and plotting scripts, as well as language refinement.

\textbf{Author contributions} 
E.P. ran the simulation grids, performed the analysis of the results and wrote the majority of the manuscript.
T.L. participated in the conceptualization of the project, supervised E.P., and contributed to the interpretation of the results, the paper writing, and detailed feedback on the draft.
J.T. ran team meetings and provided verbal feedback and guidance on this work during those meetings, gave extensive feedback on the manuscript.
H.N. developed the \texttt{AGNI} code, helped to guide the orbital migration method development, and provided general guidance, results interpretation, and detailed feedback on the draft.
M.A. provided the theoretical migration framework and guided its interpretation, contributed to supervision, and gave extensive feedback on the manuscript and figures.
T.L., H.N., E.P. and M.A. develop and maintain the \texttt{PROTEUS} framework used in this work.
A.P. provided theoretical guidance and interpretation relating to atmospheric structure and comparison to the observations, and detailed comments on the paper draft.
L.D participated in team meetings and provided editorial suggestions.
N.W. performed some of the original data reductions that helped to contextualize the results. 
M.P. and B.P. contributed to the discussion of the results.
A.M. used the modeling presented here to inform high mean molecular weight GCM simulations.
S.B. highlighted the possibility of a CO$_2$ dominated atmosphere during team meetings, in accordance with results from the phase curve paper.
All co-authors have read and provided feedback on the manuscript.

\end{acknowledgments}

\appendix
\twocolumngrid
\section{Late inward migration scenario B}

In this Appendix, we examine the late inward migration scenario B in more detail, over the course of the planet's evolution.

\begin{figure}
    \centering
    \includegraphics[width=1.0\columnwidth]{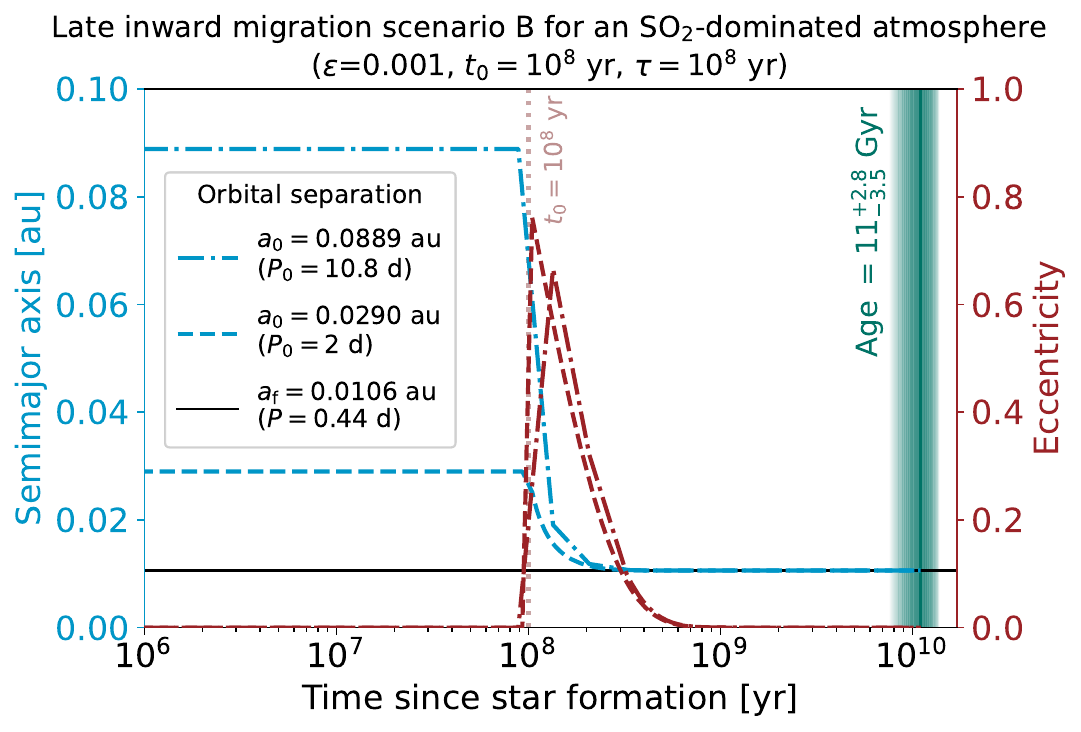}
    \caption{Evolution of the semimajor axis (blue, left axis) and eccentricity
    (red, right axis) as a function of time for the late inward migration
    scenario B, shown for two SO$_2$-dominated cases consistent with the
    present-day observations. Two initial orbital distances are tested:
    $a_0=0.0290$\,au (dashed line) and $a_0=0.0889$\,au (dashed dotted line). The present-day orbit of TOI-561\,b is indicated by the black solid line.}
    \label{plot appendix sma ecc evolution}
\end{figure}

We present the evolution of the orbital parameters introduced in Sect.\,\ref{section scenario a and b in methods}. Fig.\,\ref{plot appendix sma ecc evolution} shows the semimajor axis (blue) and eccentricity (red) as functions of time for the two initial orbital distances, $a_0=0.0290$\,au and $a_0=0.0889$\,au, for the SO$_2$-dominated cases that satisfy the present-day observational constraints. In both cases, migration is initiated at $t_0=10^{8}$\,yr and requires a high initial eccentricity ($e_0\approx0.80$) to deliver TOI-561\,b onto its observed orbit (black solid line). Because the orbital quantities are sampled on the adaptive interior--atmosphere time step, the first output past $t_0$ falls slightly after $10^{8}$\,yr (at $\approx1.3\times10^{8}$\,yr for the widest orbit at $a_0=0.0889$\,au, dash-dotted line), so the eccentricity maxima in Fig.\,\ref{plot appendix sma ecc evolution} appear shifted and mismatched. This is an output-cadence effect, not a physical inconsistency.

\bibliography{references}{}
\bibliographystyle{aasjournalv7}

\end{document}